\documentclass{aa}
\usepackage[dvips]{color}
\usepackage{graphicx}
\usepackage{longtable}
\usepackage{rotating}
\usepackage{txfonts}

\begin{document}

\title{Ground-based CCD astrometry with wide field
imagers\thanks{Based on data acquired using the Large Binocular
Telescope (LBT) at Mt. Graham, Arizona, under the Commissioning of the
Large Binocular Blue Camera.  The LBT is an international
collaboration among institutions in the United States, Italy and
Germany. LBT Corporation partners are: The University of Arizona on
behalf of the Arizona university system; Istituto Nazionale di
Astrofisica, Italy; LBT Beteiligungsgesellschaft, Germany,
representing the Max-Planck Society, the Astrophysical Institute
Potsdam, and Heidelberg University; The Ohio State University, and The
Research Corporation, on behalf of The University of Notre Dame,
University of Minnesota and University of Virginia.}}
\subtitle{IV. An improved Geometric Distortion Correction for the Blue
prime-focus Camera at the LBT.}

\author{Bellini, A.\inst{1,}\inst{2,}$\!\!$
\thanks{Visiting PhD Student at STScI under the {\it ``2008 graduate
research assistantship''} program.}
\and Bedin, L.~R.\inst{2}}

\offprints{bellini@stsci.edu}  

\institute{ Dipartimento di Astronomia, Universit\`a di Padova, Vicolo
dell'Osservatorio 3, I-35122 Padova, Italy, EU
\\\email{andrea.bellini@unipd.it}
\and Space Telescope Science Institute, 3700 San Martin Drive,
Baltimore, MD 21218, USA \\ \email{[bellini;bedin]@stsci.edu}}

\date{Received 1 December 2009 / Accepted 19 April 2010}

\abstract{High precision astrometry requires an accurate geometric
distortion solution. In this work, we present an average correction
for the Blue Camera of the Large Binocular Telescope which enables a
relative astrometric precision of $\sim$15 mas for the $B_{\rm
Bessel}$ and $V_{\rm Bessel}$ broad-band filters.  The result of this
effort is used in two companion papers:\ the first to measure the
absolute proper motion of the open cluster M~67 with respect to the
background galaxies;\ the second to decontaminate the color-magnitude
diagram 
of M~67 from field objects, enabling the study of the end of its white
dwarf cooling sequence.  Many other applications might find this
distortion correction useful.}


\maketitle

\section{Introduction}
\label{sec1}

Modern wide field imagers (WFI) equipped with CCD detectors began
their operations at the end of the last century, however -- after more
than 10 years -- their astrometric potential still remains somehow
unexploited (see Anderson et al.\ \cite{anderson06}, hereafter
Paper~I).  It is particularly timely to begin exploring their full
potential now that WFI start to appear also at the focus of the
largest available 8m-class telescopes.

The present work goes in this direction, presenting a correction for
the geometric distortion (GD) of the Blue prime-focus Large Binocular
Camera (LBC), at the Large Binocular Telescope (LBT).  Unlike in
Paper~I, in which we corrected the GD of the WFI at the focus of the
2.2m MPI/ESO telescope (WFI@2.2m) with a look-up table of corrections,
for the LBC@LBT we will adopt the same technique described in Anderson
\& King (\cite{AK03}, hereafter AK03), and successfully applied to the
new Wide field Camera 3/UV-Optical channel on board the {\it Hubble
Space Telescope} (Bellini \& Bedin \cite{bb09}, hereafter BB09).

This article is organized as follows: Section 2 briefly describes the
telescope/camera set up; Section 3 presents the data set used.  In
section 4, we describe the steps which allowed us to obtain a solution
of the GD, for each detector separately, while in Section 5 we
presents a (less accurate) inter-chip solution.  Distortion stability
is analyzed in Section 6, and a final Section summarizes our results.

\begin{figure*}[ht!]
\centering
\includegraphics[height=10.2cm]{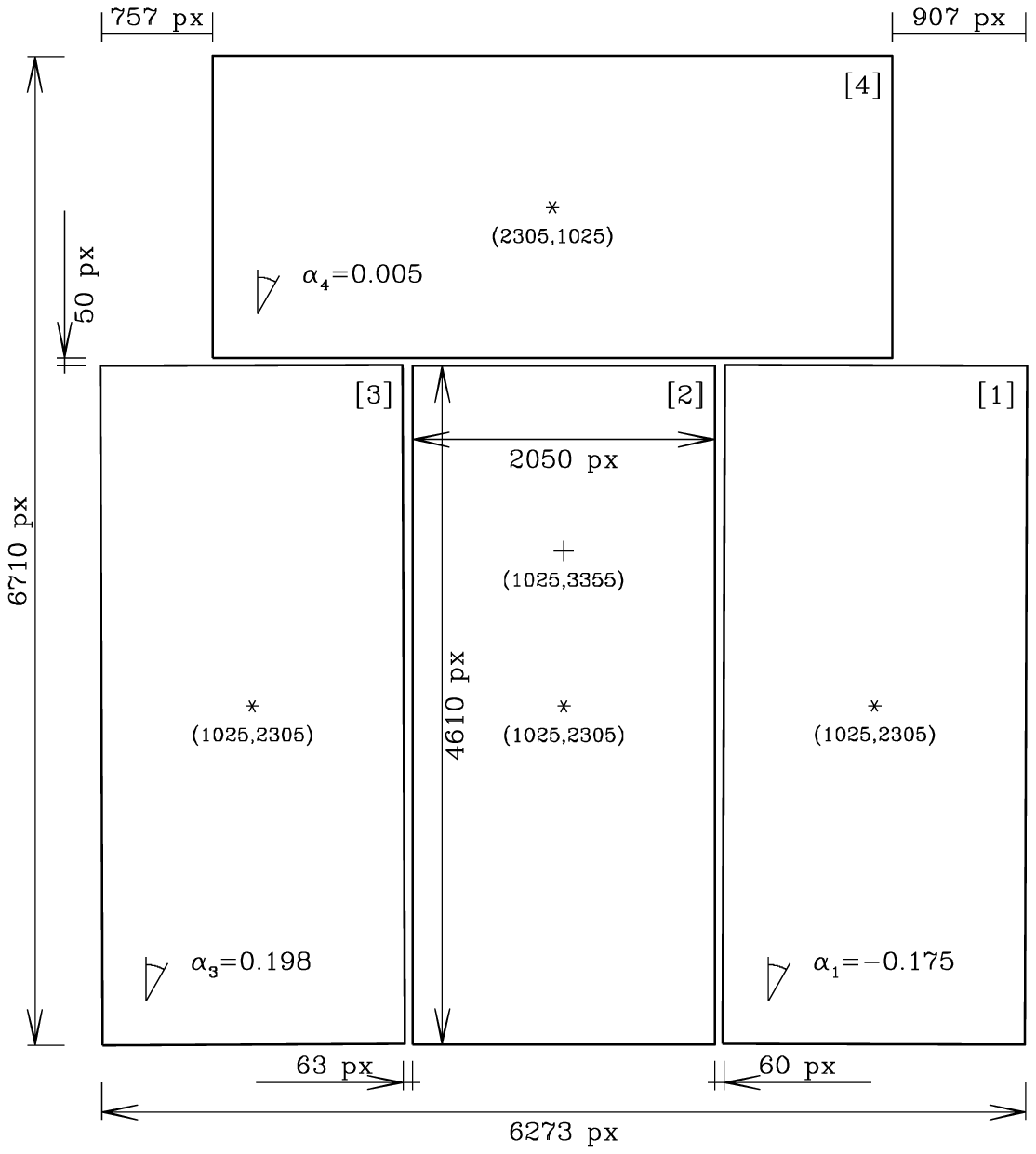}
\includegraphics[height=10.0cm]{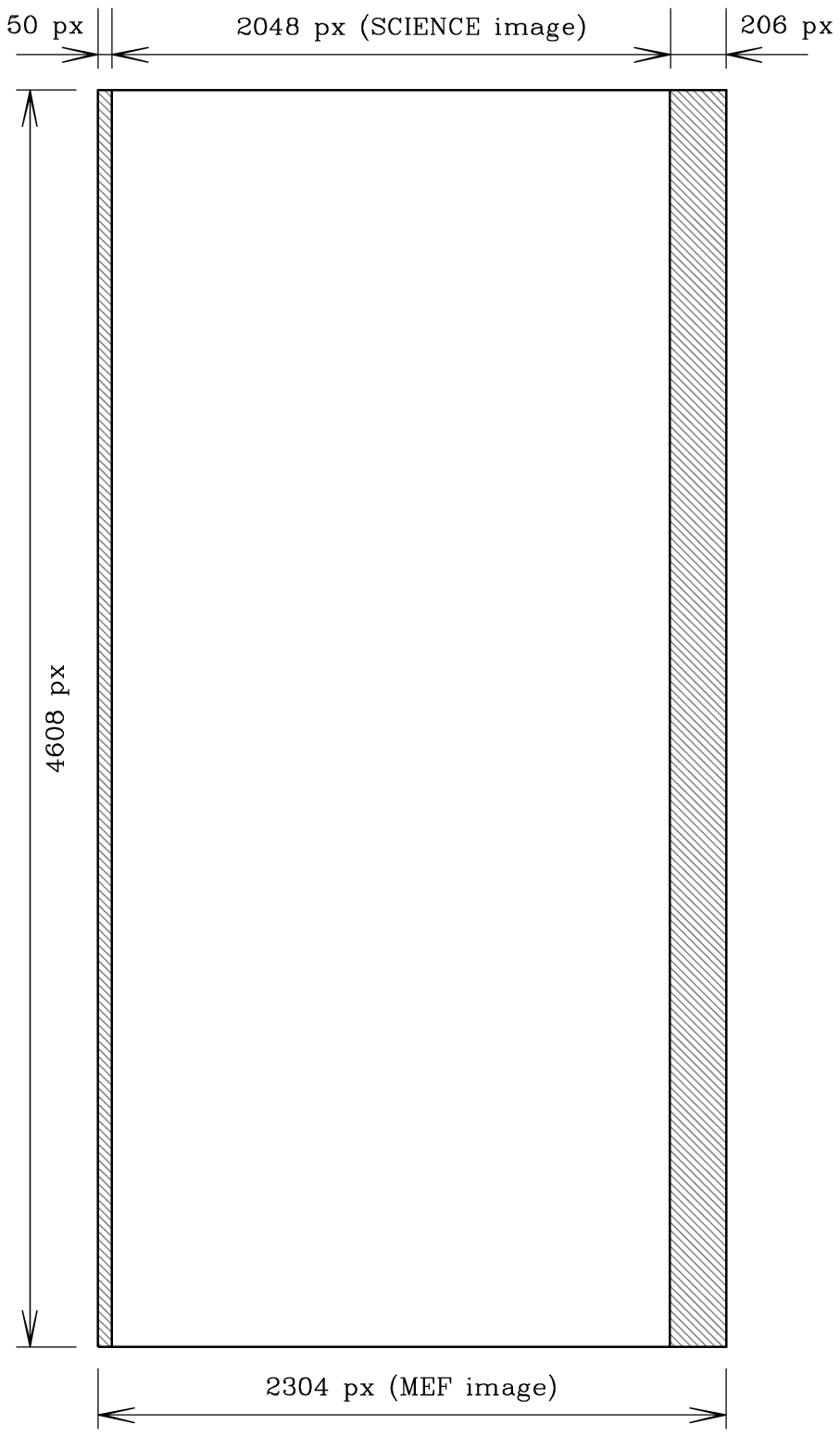}\\
\caption{(\textit{Left}): LBC-Blue mosaic layout; ``$\ast$'' marks the
center for each chip (see Sect.\ \ref{sec:3} for the operative
definition of centers and chips), while ``+'' marks the here defined
center of the mosaic.  (\textit{Right}): Each MEF file consists of 4
images. Each image is composed by one scientific region and two
overscan regions, covering the first 50 and the last 206 pixel columns
(shaded).}
\label{fig1}
\end{figure*}

\section{The Large Binocular Camera Blue}
\label{sec:2}

The LBT is a large optical/infrared telescope that utilizes two
mirrors, each having a diameter of 8.4
meters\footnote{\texttt{www.lbt.it};
\texttt{medusa.as.arizona.edu/lbto/}.}.  The focal ratio of the LBT
primary mirrors ($F$/1.14) and its large diameter are factors that
require a careful development of the corrector for a prime-focus
camera.  The blue channel of the LBC (LBC-Blue) is mounted at the
prime focus of the first LBT unit.  The corrector, consisting of three
lenses, is designed to correct spherical aberration, coma, and field
curvature, according to the design by Wynne (\cite{wynne96}). The last
two of these three lenses are sub-divided in two elements each, with
the last one being the window of the cryostat (Ragazzoni et al.\
\cite{ragazzoni00}, \cite{ragazzoni06}; Giallongo et al.\
\cite{giallongo08}).  The final LBC-Blue focal-ratio is $F$/1.46.

The LBC-Blue employs an array of four 16-bit e2v 42-90
(2048$\times$4608) chips, with a reference pixel-scale of
$0\farcs2297\,{\rm pix}^{-1}$ (this work), providing a total Field of
View (FoV) of $\sim$$24^\prime$$\times$$25^\prime$.  The four chips
are mounted on the focal plane in such a way as to maximize the
symmetry of the field, with three chips contiguous longside, and the
fourth one rotated 90 degrees anti-clockwise, and centered above the
others.  The LBC-Blue layout is shown on the left hand of
Fig.~\ref{fig1}.  Row estimates of the intra-chip gaps are expressed
as the nearest integer pixel. Numbers between square brackets are chip
identification numbers, as read from the raw Multi Extension Fits
(MEF) file.  Average rotation angles are given with respect to chip \#
2, chosen as reference (we will see in Section~\ref{sec:8} how to
bring positions from each chip into a common corrected meta-chip
system).  On the right hand of Fig.~\ref{fig1} we show, in units of
raw pixel coordinates, the dimensions of each chip, which consists of
the scientific image in between two overscan regions (shaded areas
Fig.~\ref{fig1}, which cover the first 50 and the last 206 pixel
columns).

During the optical design phase, GD (of pin-cushion type) was not
considered as an aberration, since it may be corrected at
post-processing stages.  The GD is found to be always below the 1.75\%
level (Giallongo et al.\ \cite{giallongo08}).  This is translated in
offsets as large as 50 pixels ($\sim$11 arcsec) from corner to corner
of the LBC-Blue FoV.  Obviously, the correction of such a large GD is
of fundamental importance for high precision astrometric measurements.
Note that in the following, with the term ``geometric distortion'' we
are lumping together several effects: the optical field-angle
distortion introduced by camera optics, light-path deviations caused
by the filters (in this case $B_{\rm Bessel}$ and $V_{\rm
Bessel}$), non-flat CCDs, alignment errors of CCDs on the focal
plane, etc.

\begin{table}[t!]
\centering
\caption{Log of M~67 data used in this work.}
\label{tab1}
\scriptsize{
\begin{tabular}{ccccc}
\hline \hline & & & & \\ {\bf Date}&{\bf Filter}&{\bf
\#Images$\times$Exp. time}&{\bf Airmass}& {\bf Image Quality}\\ & &
(s) & $(\sec z)$ &(arcsec)\\ & & & & \\ \hline & & & & \\ Feb 22,
2007&$V_{\rm Bessel}$&
$\!\!\!1\times15$,$1\times330$,$17\times110\!\!\!$&
1.07-1.13&0.62-1.31\\ &&&&\\ Feb 27, 2007& $V_{\rm
Bessel}$&$25\times100$&1.07-1.10&0.84-1.26\\ &&&&\\ Mar 16, 2007&
$B_{\rm Bessel}$&$25\times180$&1.07-1.14&0.84-1.08\\ & & & & \\ \hline
\end{tabular}}
\end{table}

\begin{figure*}[t!]
\centering
\includegraphics[width=9cm]{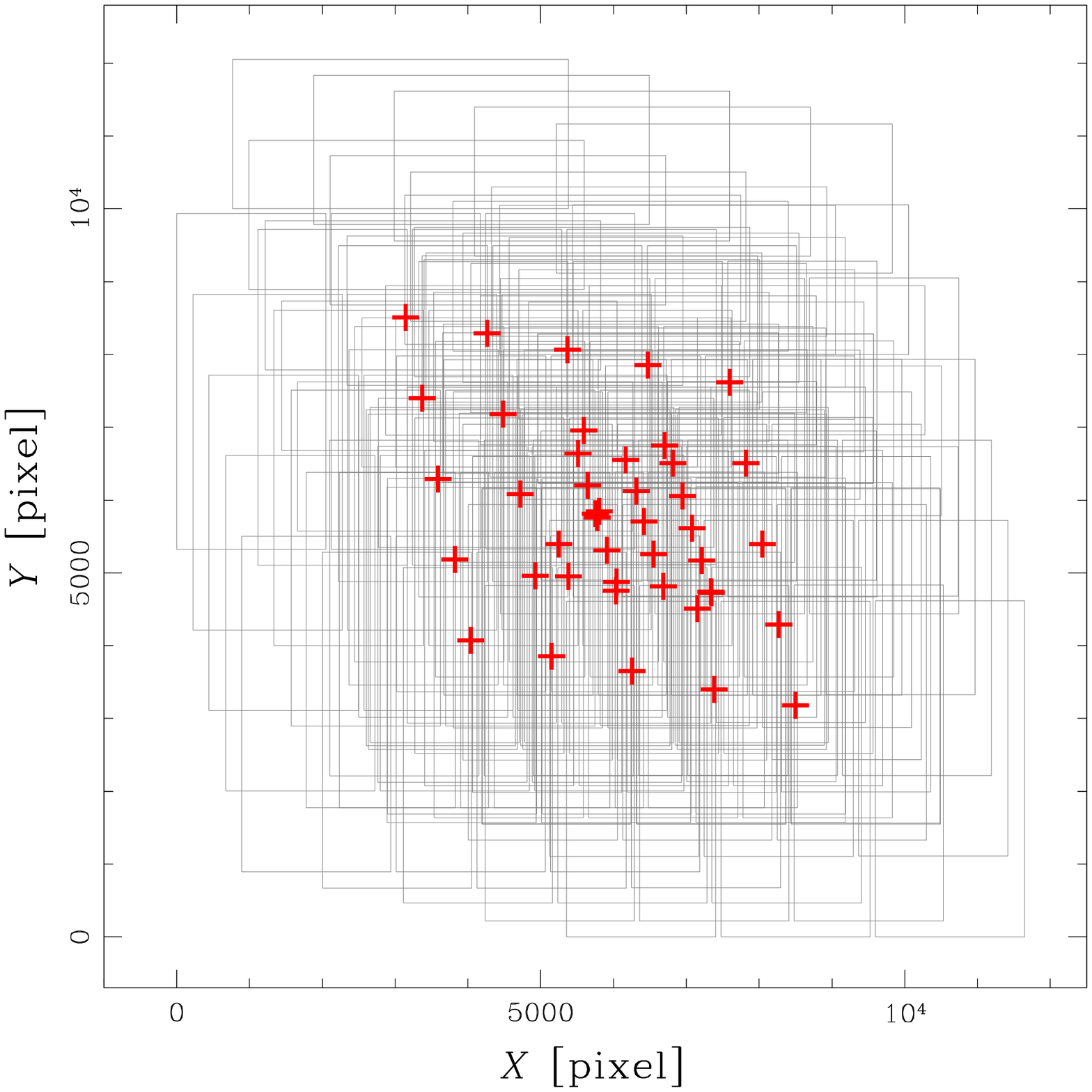}\hskip 5mm
\includegraphics[width=7cm]{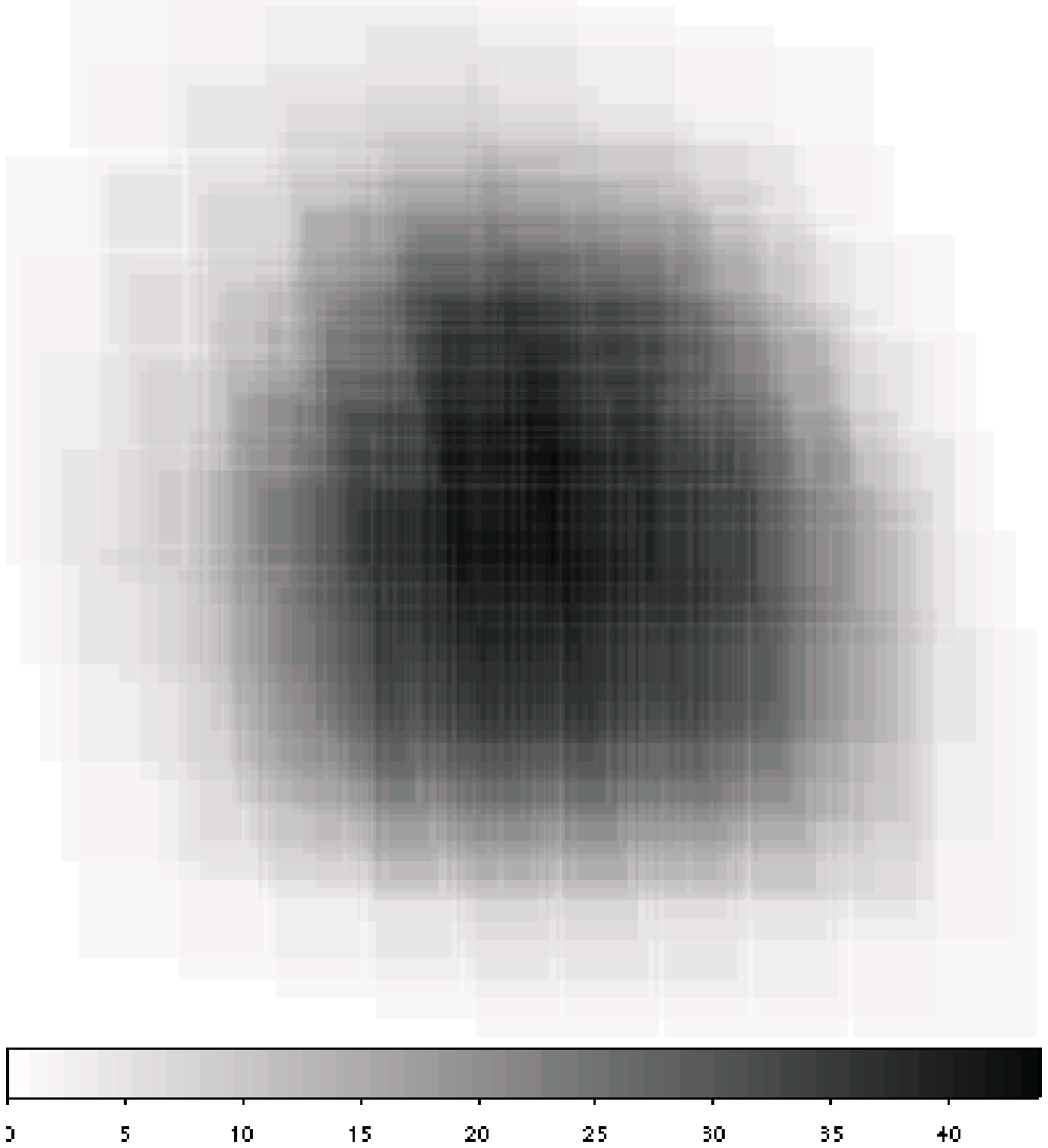}\\
\caption{\textit{(Left)}: Dither pattern, in pixel units, of the $V$
images used to solve for the geometric distortion.  We used both small
and large dithers to adequately sample the GD.  Red crosses mark the
center of LBC-Blue mosaic, (1025, 3355) position in the coordinate
system of chip \# [2], as defined in Fig.\ \ref{fig1}).  The original
pattern was designed to be a rhomboidal array of 5$\times$5 pointings
(see text). Unfortunately some of the images with the small dither
pattern were not usable.  \textit{(Right):} Depth-of-coverage map for
the same images.}
\label{fig2}
\end{figure*}

Raw data images are contained in a single MEF file with four
extensions, one for each chip, constituted by 2304$\times$4608 pixels,
containing overscan regions.  The scientific area of these chips is
located within pixel (51, 1) and pixel (2098, 4608)
(i.e. 2K$\times$4.5K pixels, see right panel of Fig.\ \ref{fig1}).
For reasons of convenience, we added an extra pixel (flagged at a
value of $-$475) to define the borders of the 2048$\times$4608 pixel
scientific regions, and we will deal exclusively with 2050$\times$4610
arrays ({\it work-images}).  Again for reasons of convenience, since
chip \# [4] is stored along the same physical dimensions as the other
three in the raw MEF file, we decided to rotate it by 90 degrees
anti-clockwise.

Hereafter, when referring to $x$ and $y$ positions, we will refer to
the raw pixel coordinates measured on these work-images, and -- unless
otherwise specified -- we will refer to the work-image of the chip \#
[$k$], simply as [$k$].  Transformation equations to convert from the
raw pixel coordinates of the archive MEF file $(x^{\rm MEF}_{k},y^{\rm
MEF}_{k})$ to the pixel coordinates of the work-images $(x,y)$ are as
follows:
$$
k=1,2,3: ~ \left\{
\begin{array}{rcl}
\displaystyle x&\!\!\!=\!\!\!&
x^{\rm MEF}_{k}-49\\
\displaystyle y&\!\!\!=\!\!\!&y^{\rm MEF}_{k}+1\\
\end{array}\right.\\
$$
$$
 ~ k=4: ~ 
\left\{
\begin{array}{rcl}
\displaystyle x&\!\!\!=\!\!\!&4610-y^{\rm MEF}_{k}\\
\displaystyle y&\!\!\!=\!\!\!&x^{\rm MEF}_{k}-49.\\
\end{array}\right.$$

\noindent
[For clarity, every LBC-Blue image is a MEF file, from which we define
4 {\it work-images}. Moreover, we will treat every chip of each image
independently.]

\section{The data-set}
\label{sec:3}

During LBT science-demonstration time, between February and March
2007, we obtained (under the Italian guaranteed time) about four hours
to observe the old, metal-rich open cluster M~67 ($\alpha=08^{\rm
h}51^{\rm m}23^{\rm s}.3$, $\delta=+11^{\circ}49\arcmin02\arcsec$,
J2000.0, Yadav et al.\ \cite{yadav08}, hereafter Paper~II).  The aim
of the project is to reach the end of the DA white dwarf (WD) cooling
sequence (CS) in the two filters $V_{\rm Bessel}$ and $B_{\rm Bessel}$
(hereafter simply $V$ and $B$).  In addition, we want to compute
proper motions for a sample of objects in the field by combining these
LBC@LBT exposures with archival images collected 10 years before at
the Canada France Hawaii Telescope (CFHT).  The pure sample of WD
members will serve to better understand the physical processes that
rule the WD cooling in metal-rich clusters.  A necessary first step to
get accurate proper motions is to solve the GD for the LBC-Blue.  The
results of the investigation on the WD CS of M~67, and its absolute
proper motion, are presented in two companion papers (Bellini et al.\
\cite{bellini10a}, \cite{bellini10b}); here we will focus on the GD of
LBC-Blue, providing a solution that might be useful to a broader
community of LBC-Blue users.

The observing strategy had to arrange both the scientific goals of the
project and the need to solve for the geometric distortion.
As an educated guess, the adopted procedure to solve for the geometric
distortion is the auto-calibration described in great detail in
Paper~I, which still represents the state of the art in ground-based
CCD astrometry with wide-field imagers.

With the idea to map the same patch of the sky in different locations
on the same chip, as well as on different chips, we chose a particular
pointing set up, constituted by an array of 5$\times$5 observations,
dithered in such a way that a star never falls two times on the same
gap between the chips.  All 25 exposures of a given dither sequence
were executed consecutively.  The 5$\times$5 dither pattern is
repeated adopting small ($\sim$100$\arcsec$) and large
($\sim$200$\arcsec$) steps in filter $V$, and only small steps in the
$B$ filter.  Figure~\ref{fig2} shows the dither pattern and the
depth-of-coverage map for all our $V$ exposures. Table~\ref{tab1}
gives the log of observations for both $B$ and $V$ exposures.  All the
images were collected in service mode.

Unfortunately, not all the exposures met the desired specifications of
our proposal (dark-night conditions and seeing better than
$0\farcs8$).  In particular, all the $V$ images with large dithers are
affected by anomalously high background values (up to $\sim$20$\,$000
counts for a 100 s exposure, thus limiting us at the faint
magnitudes).  Moreover, 6 out of the 25 $V$ images taken with small
dithers have an image quality well above $1\farcs5$ (probably related
to guide-star system problems).  These images are of no use for our
purpose, and were not considered in the present study.

Our GD solution will be first obtained for the $V$ filter images, and
later tested, and eventually re-derived, for the $B$ filter ones.  To
measure star positions and fluxes, we developed a reduction method
that is mostly based on the software {\sf img2xym\_WFI} (Paper~I).
This new software ({\sf img2xym\_LBC}) similarly generates a list of
positions, fluxes, and a quality of the PSF-fit values (see Anderson
et al. \cite{anderson08}) for each of the measured objects in each of
the four chips.  Details of the PSF-fitting software {\sf
img2xym\_LBC} and the final M~67 astro-photometric catalog will be
presented in a subsequent paper of this series (which will also deal
with photometric zero point variations and PSF variability).

\section{Auto-calibration}
\label{sec4}

The most straightforward way to solve for the GD would be to observe a
field where there is a prior knowledge of the positions of all the
stars in a distortion-free reference frame.  [A distortion-free
reference frame is a system that can be transformed into any another
distortion-free frame by means of {\it conformal
transformations}\footnote{A conformal transformation between two
catalogs of positions is a four-parameter linear transformation,
specifically: rigid shifts in the two coordinates, one rotation, and
one change of scale, i.e. the shape is preserved.}.]  GD would then
show itself immediately as the residuals between the observed relative
positions of stars and the ones predicted by the distortion-free frame
(on the basis of a conformal transformation).  Unfortunately, such an
``astrometric flat-field'' with the right magnitude interval, source
density, and accuracy, is difficult to find and astronomers are often
left with the only option of auto-calibration.

The basic principle of auto-calibration is to observe the same stars
in as many different locations on the detector as possible, and to
compute their average positions once they are transformed onto a
common reference frame\footnote{
We want to make clear that we had at our disposal only $\sim$4 hours
of telescope time during the science-demonstration time, to be used
both for the science and the calibration project. With the minimum
exposure time needed to have a good signal to noise ratio for the
target stars ($\sim$100 s), and taking into account overheads for the
necessarily large dithers for GD correction, the optimal solution was
to observe 25 dithered exposures with the aim of calibrating the LBC
distortion.
}.  Ideally, a star should be observed from corner to corner in the
FoV. This means that the total dither has to be as large as the FoV
itself (see Fig.\ \ref{fig2}).

If the observations are taken with a symmetric dither pattern, the
systematic errors will have a random amplitude, and the stars'
averaged position will provide a better approximation of their true
position in a distortion-free frame (the master frame).  This master
frame -- as defined by the averaged position of the sources in the
FoV -- will then serve as a first guess for the construction of an
astrometric flat-field, which in turn can be used (as we will see in
detail below) to compute star-position residuals (hereafter simply
residuals), necessary to obtain a first estimate of the GD for each
chip.  Single chips are then individually corrected with these
preliminary GD solutions (one for each chip) and the procedure of
deriving the master frame is repeated.  With the new-derived master
frame, new (generally smaller) residuals are computed, and the
procedure is iteratively repeated until convergence is reached (see
below).

The overall distortion of LBC-Blue is large enough ($\sim$50 pixels)
that -- to facilitate the cross-correlation of positions of objects
observed in different locations on the detector -- it becomes very
convenient to perform a preliminary (although crude) correction.

As a first guess for the master frame, we used the best astrometric
flat-field available in the literature for the M~67 field: the
astro-photometric catalog recently published in Paper~II.  This
catalog was obtained with images taken with the WFI@2.2m; it is deeper
with respect to other wide-field catalogs (i.e., UCAC2, USNO-A2, and
2MASS), has $V$ photometry, and its global astrometric accuracy is of
the order of $\sim$50 mas.  Nevertheless, this catalog is far from
ideal; even the faintest -- poorly measured -- stars of Paper~II are
close to saturation in our LBC-Blue images, and the total number of
usable (even if saturated) objects was never above $\sim$250 per chip
(among which less than $\sim$40 per chip were unsaturated).  We also
chose to re-scale the pixel coordinates of the Paper~II catalog (with
an assumed WFI@2.2m pixel-scale of 238 mas, Paper~I) to the average
pixel-scale of LBC-Blue, adopting for it the median value of $225.4$
mas/pixel (as derived by Giallongo et al.\ \cite{giallongo08}).  Since
the scale is a free parameter in deriving GD correction, choosing a
particular scale value will not invalidate the solution itself.  Later
we will derive the average scale of [2] in its central pixel
(1025,2305), and we will determine the absolute value of our
master-frame plate scale by comparison with objects in the Digital Sky
Survey, and study the average inter-chip scale variations with time
and conditions.

Once this first-guess solution is obtained, it is easier to
cross-correlate the star catalogs from each LBC-Blue work-image with
respect to a common reference frame, in order to perform the
auto-calibration procedure, as described in detail in the following
subsections.

\subsection{Deriving a self-consistent solution}
\label{sec4.1}

We closely followed the auto-calibration procedures described in
detail -- and used with success -- in AK03 to derive the GD correction
for each of the four detectors of WFPC2.  The auto-calibration method
consists of two steps:\ 1) deriving the master frame, and 2) solving
for the GD for each chip, individually.  These two steps are then
repeated interactively, until both the geometric distortion solutions
and the positions in the master-list converge.

\begin{figure}[t!]
\centering \includegraphics[width=8.8cm]{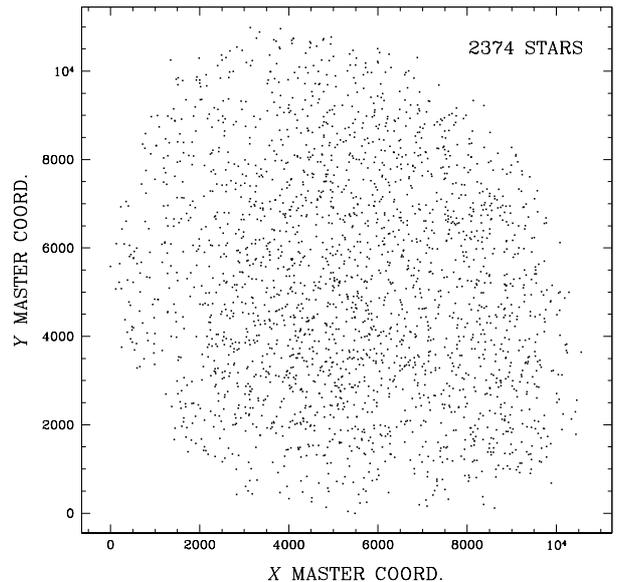}
\caption{The master star-list map.}
\label{fig:master}
\end{figure}

\subsubsection{The master-list}
\label{sec4.2}

As aforementioned, only during the very first iteration did we use the
Paper~II catalog as a master frame to get the preliminary best guess
of the GD for each chip.  In all the subsequent iterations, the master
frame was obtained from all the available LBC work-images (i.\ e., the
master-list is made with images taken within few days).  Conformal
transformations are used to bring star positions, as measured in each
work-image, into the reference system of the current master frame.  We
used only well-measured, unsaturated objects with a stellar profile.
The final master-list contains 2374 uniformly spread stars (see
Fig.~\ref{fig:master}), with coordinates $(X^{\rm master},Y^{\rm
master})_i$, with $i=1,\dots,2374$, that were observed, at each
iteration, in at least 3 different images. As we can see on the right
panel of Fig.\ \ref{fig2}, stars falling in the center of our FoV can
be observed up to 44 times in the $V$-filter, i.\ e.\ the maximum
overlap among the $V$ exposures.  We have at most 25 observations for
a given star in the case of the $B$-filter exposures.

\subsubsection{Modeling the geometric distortion}
\label{sec4.3}

As in AK03 and in BB09, we represent our solution with two third-order
polynomials.  Indeed, we found that with two third-order polynomials
our final GD correction has a precision level of $\sim$0.04 pixel in
each coordinate ($\sim$10 mas), and higher orders were unnecessary,
with this precision level (as we will see) being well within the
instrument stability. We performed tests with fourth- and fifth-order
polynomials, obtaining comparable results in term of GD-solution
accuracy, but at the expense of using a large number of degrees of
freedom in modeling the GD solution.

Having an independent solution for each chip, rather than one that
uses a common center of the distortion for the whole FoV, allows a
better handle on individual detector effects, such as a different
relative tilt of the chip surfaces, etc.  We chose a pixel close to
the physical center of each chip as reference position, with respect
to solve for the GD, regardless of its relative position with respect
to the principal axes of the optical system.  The adopted centers of
our solution are the locations
$(x_\circ,y_\circ)_{k=1,2,3}=(1025,2305)$ for chips [1], [2], [3], and
the $(x_\circ,y_\circ)_{4}=(2305,1025)$ for chip [4], all in the raw
pixel coordinates of the work-images.

For each $i$-star in each $k$-chip of each $j$-MEF file, the
distortion corrected position $(x_{i,j,k}^{\rm corr},y_{i,j,k}^{\rm
corr})$ is the observed position plus the distortion correction
$(\delta x_{i,j,k},\delta y_{i,j,k})$:
$$
\left\{
\begin{array}{c}
\displaystyle x_{i,j,k}^{\rm corr}=x_{i,j,k}+\delta x_{i,j,k}(\tilde{x}_{i,j,k},\tilde{y}_{i,j,k})\\
\displaystyle y_{i,j,k}^{\rm corr}=y_{i,j,k}+\delta y_{i,j,k}(\tilde{x}_{i,j,k},\tilde{y}_{i,j,k}),\\
\end{array}
\right.
$$
where $\tilde{x}_{i,j,k}$ and $\tilde{y}_{i,j,k}$ are the normalized
positions, defined as:
$$
\left\{
\begin{array}{c}
\displaystyle\tilde{x}_{i,j,k}= \frac{x_{i,j,k}-(x_\circ)_{k}}{(x_\circ)_{k}}\\
\displaystyle\tilde{y}_{i,j,k}= \frac{y_{i,j,k}-(y_\circ)_{k}}{(y_\circ)_{k}}.\\
\end{array}
\right.
$$ 
[Normalized positions make it easier to recognize the magnitude of the
contribution given by each solution term, and their numerical
round-off.]

\begin{figure*}[t!]
\centering
\includegraphics[height=5.3cm]{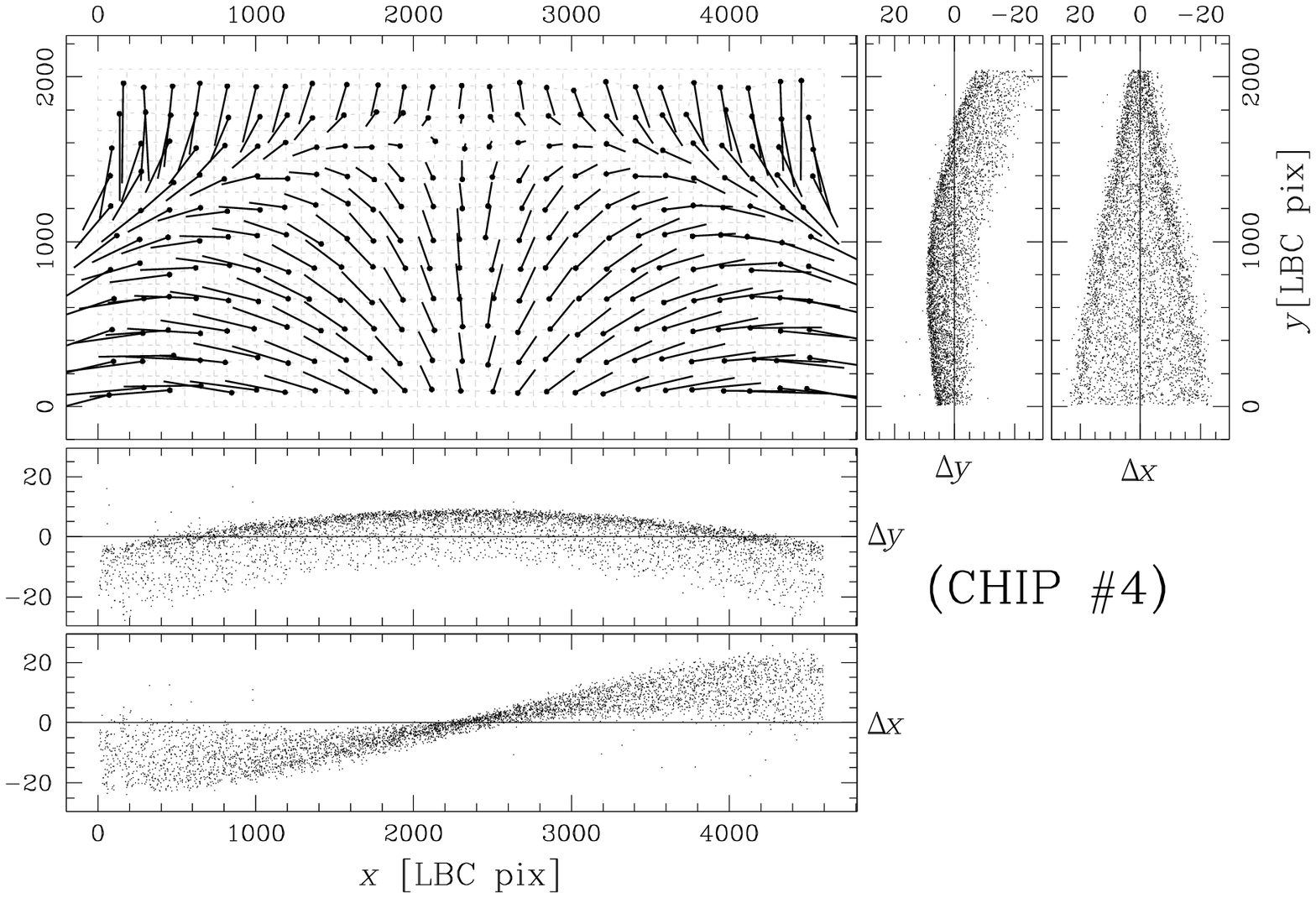}\\
\includegraphics[ width=5.3cm]{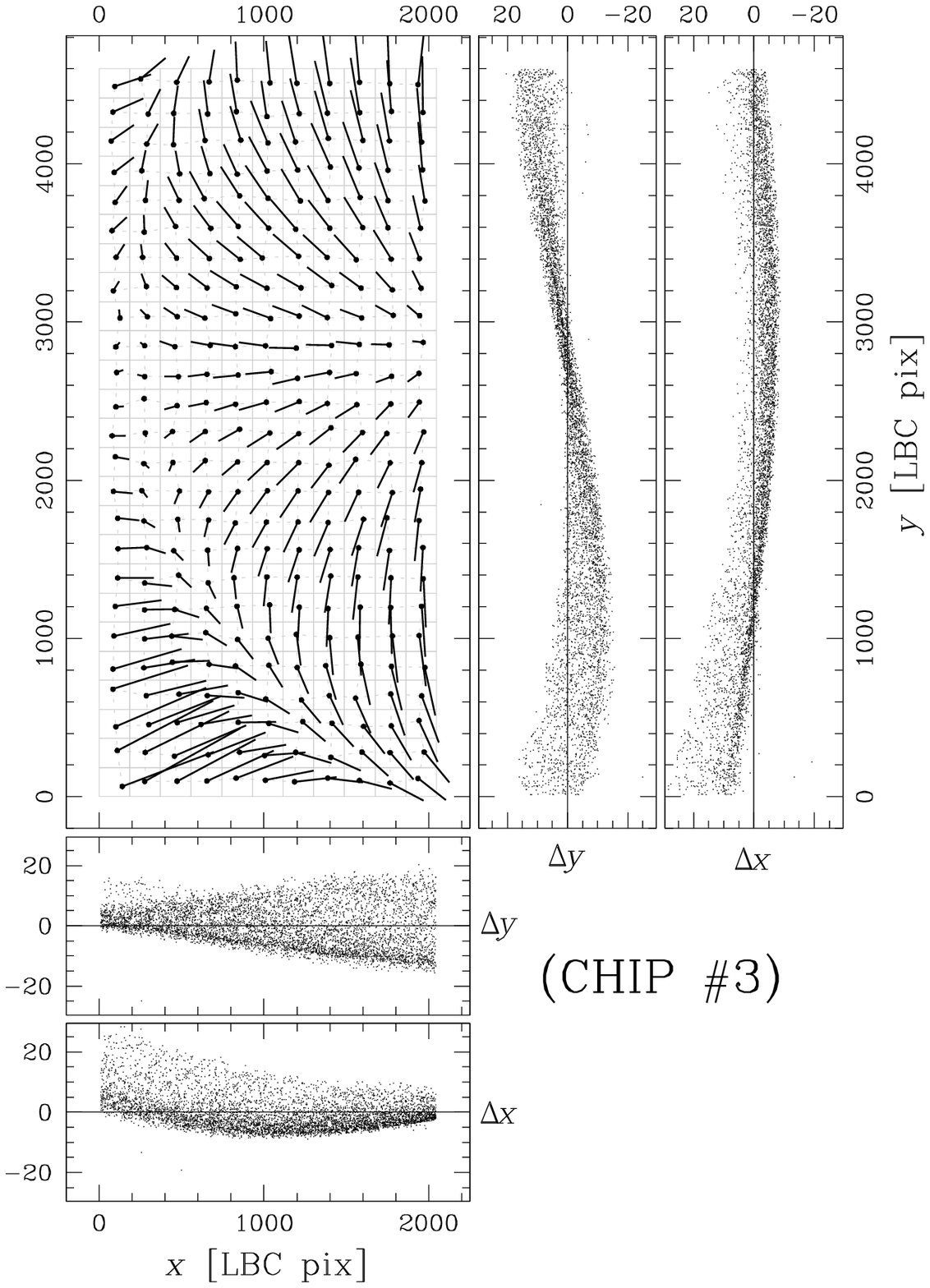}
\includegraphics[ width=5.3cm]{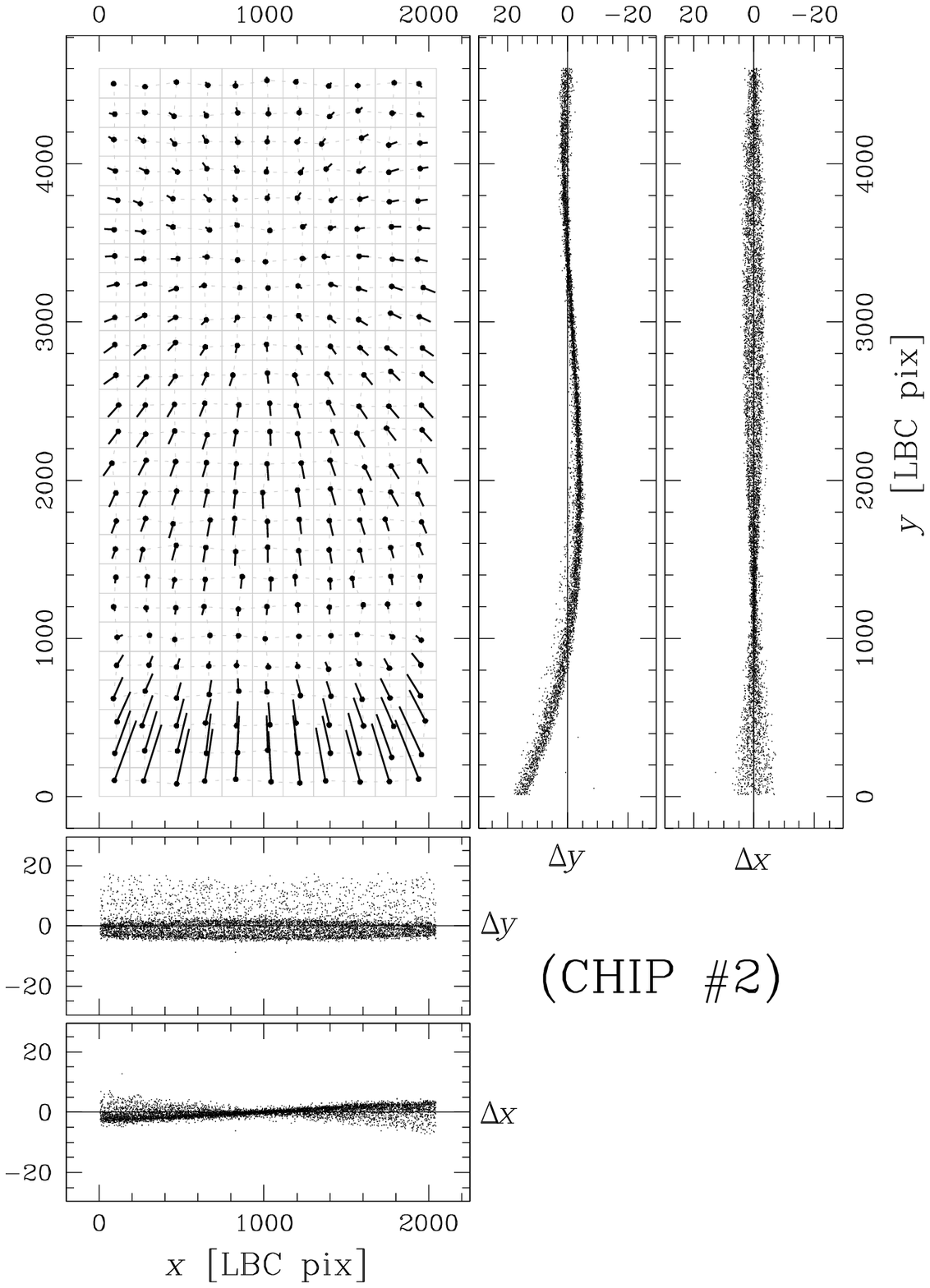}
\includegraphics[ width=5.3cm]{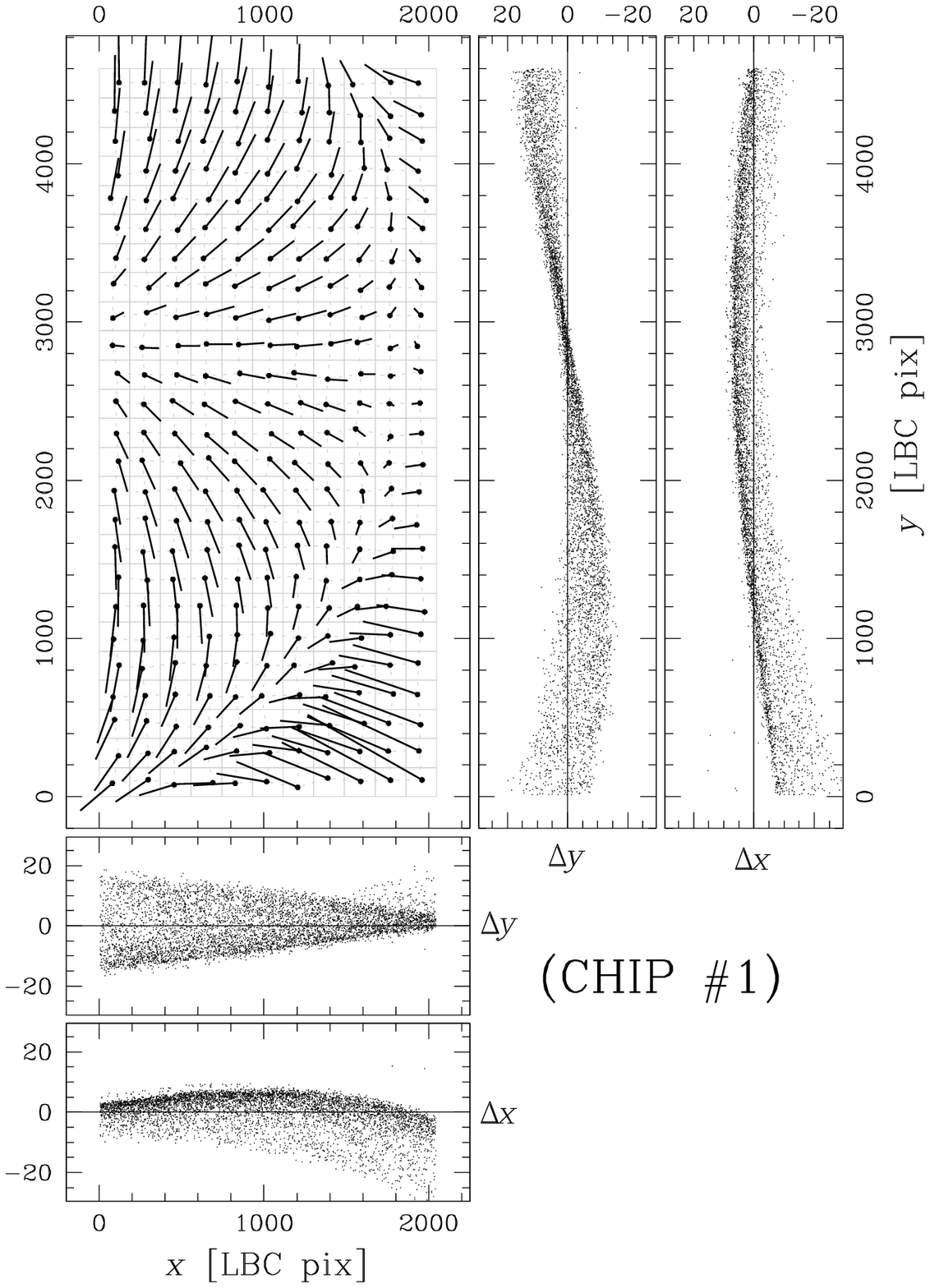}\\
\caption{Residual trends for the four chips when we use uncorrected
star positions.  The size of the residual vectors is magnified by a
factor of 25.  For each chip, we also plot the single residual trends
along the $x$ and $y$ axes. Units are expressed as LBC-Blue raw
pixels.}
\label{fig4}
\end{figure*}

\begin{figure*}[t!]
\centering
\includegraphics[height=5.3cm]{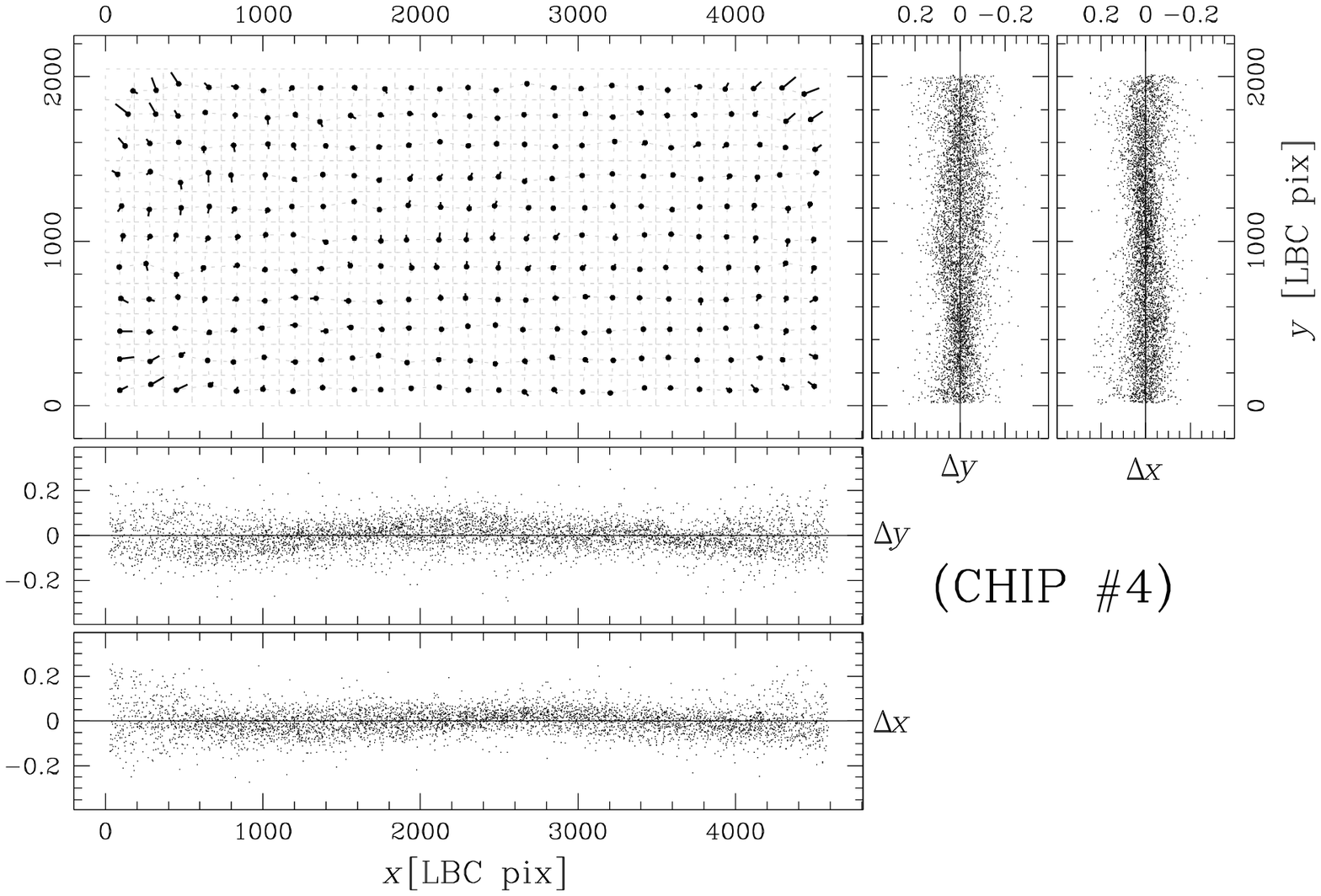}\\
\includegraphics[ width=5.3cm]{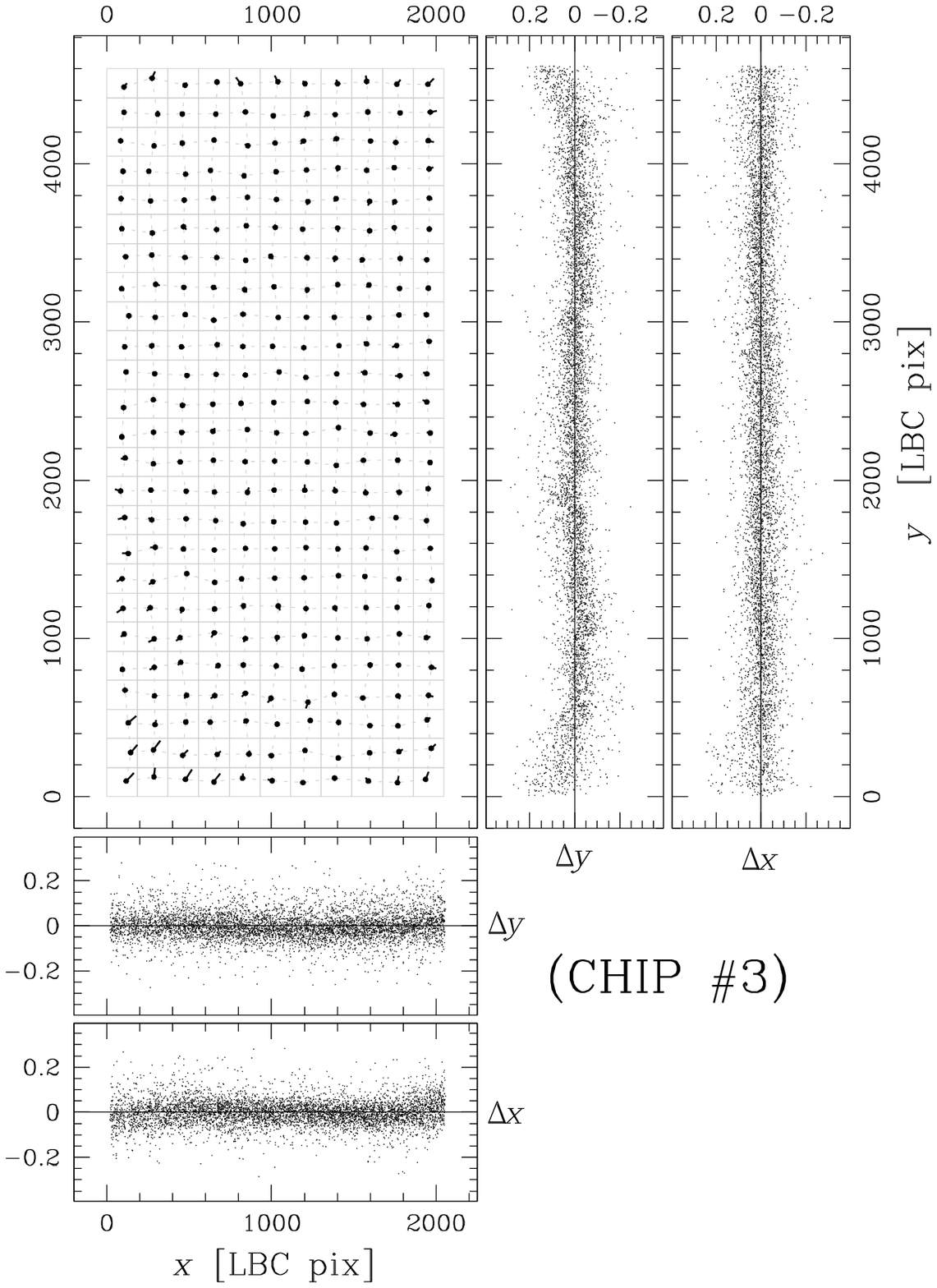}
\includegraphics[ width=5.3cm]{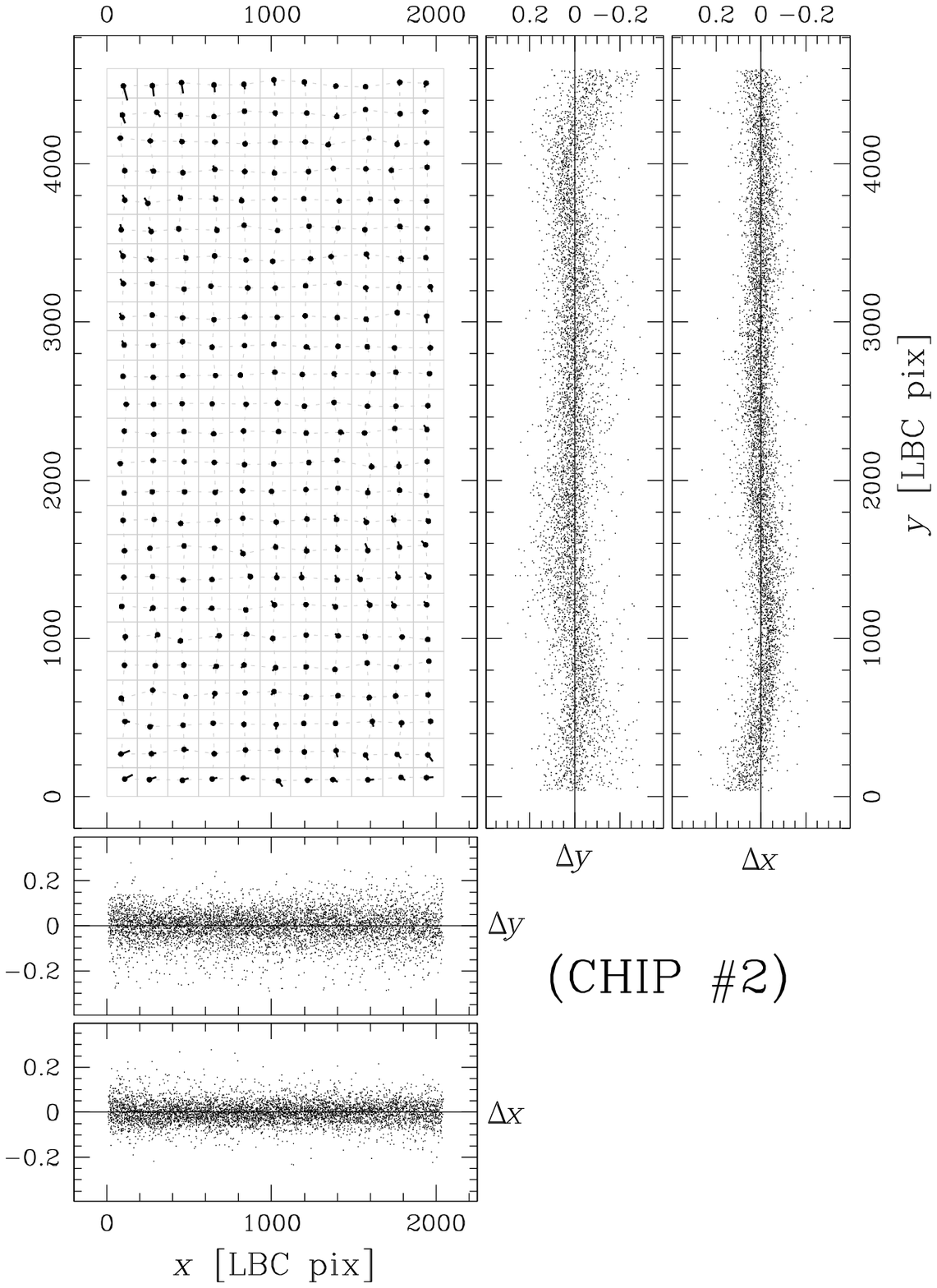}
\includegraphics[ width=5.3cm]{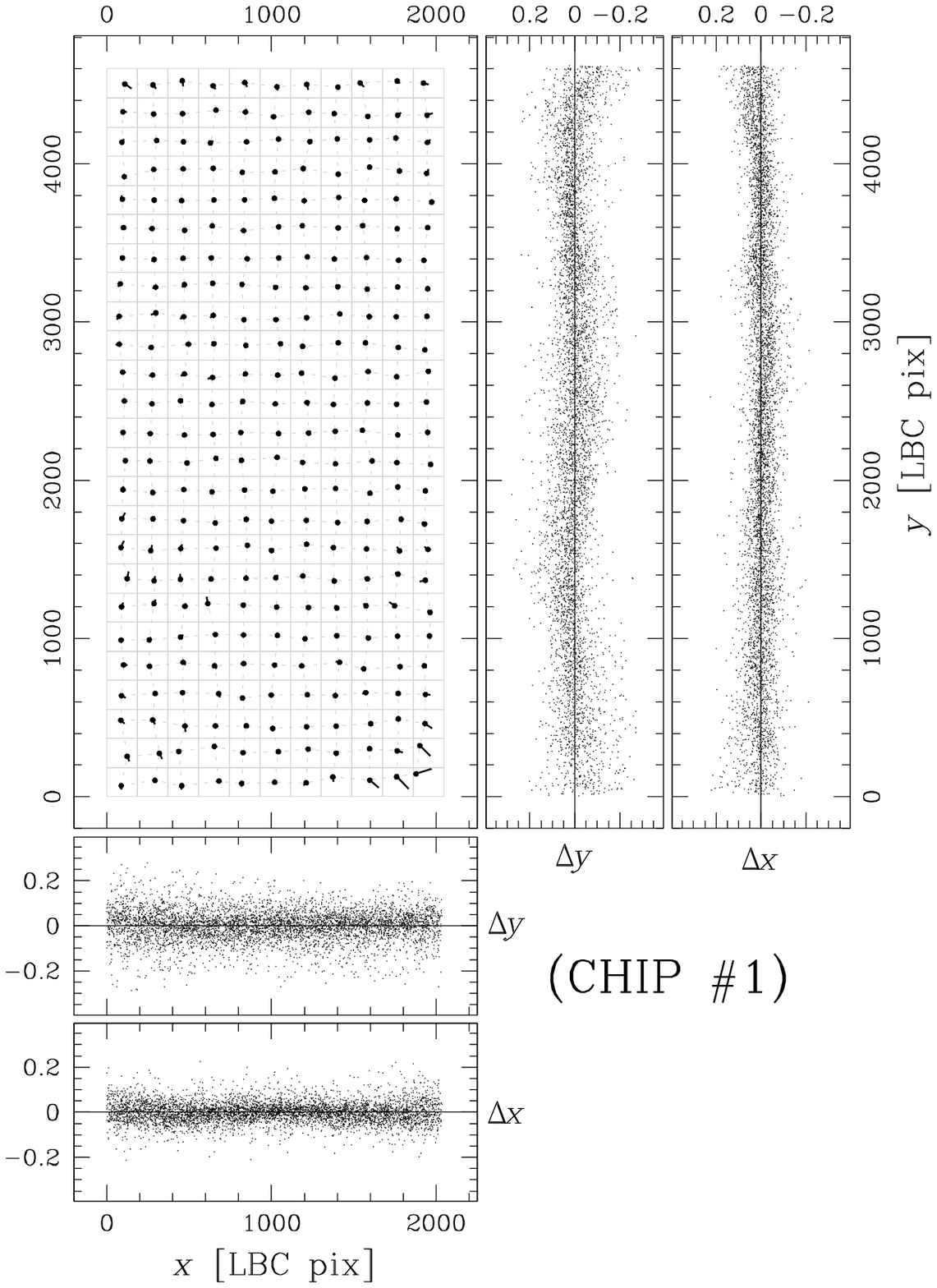}\\
\caption{Same as Fig.~\ref{fig4}, but for our corrected star
positions. The size of the residual vectors is now magnified by a
factor of 500.}
\label{fig5}
\end{figure*}

The final distortion correction, for each star in each work-image, is
given by the following two third-order polynomials (we omitted here
$i,j,k$ indexes for simplicity):
$$
\left\{
\begin{array}{rcl}
\displaystyle     \delta    x&\!\!\!=\!\!\!&a_1\tilde{x}+a_2\tilde{y}+
a_3\tilde{x}^2    +   a_4    \tilde{x}\tilde{y}+    a_5\tilde{y}^2   +
a_6\tilde{x}^3+a_7\tilde{x}^2\tilde{y}   \\
&&\phantom{a_1\tilde{x}}+  a_8\tilde{x}\tilde{y}^2  +
a_9\tilde{y}^3\\

\displaystyle \delta y&\!\!\!=\!\!\!&b_1\tilde{x}+b_2\tilde{y}+
b_3\tilde{x}^2 + b_4 \tilde{x}\tilde{y}+ b_5\tilde{y}^2 +
b_6\tilde{x}^3+b_7\tilde{x}^2\tilde{y} \\ &&\phantom{a_1\tilde{x}}+
b_8\tilde{x}\tilde{y}^2 + b_9\tilde{y}^3.\\
\end{array}
\right.
$$

Our GD solution is thus fully characterized by 18 coefficients:
$a_1,\dots, a_9$, $b_1,\dots, b_9$.  As done in AK03 and BB09, we
imposed $a_{1,k}=0$ and $a_{2,k}=0$ to constrain the solution so that,
at the center of the chip, it will have its $x$-scale equal to the one
at the location $(x_\circ,y_\circ)$, and the corrected axis $y_{\rm
corr}$ aligned with its $y$-axis at the same location.  On the other
hand, $b_{1,k}$ and $b_{2,k}$ must be free to assume whatever values
fit best, to account for differences in scale and perpendicularity of
detector's axes.  Therefore, we only have to compute 16 coefficients
(for each chip) to derive our GD solution.

\subsubsection{Building the Residuals}
\label{sec4.4}

Each $i$-star in the master frame is conformally transformed into each
$k$-work-image/$j$-file, and cross-identified with the closest source.
We indicate such transformed positions with $(X_i^{\rm
master})^{T_{j,k}}$ and $(Y_i^{\rm master})^{T_{j,k}}$.  Each of such
cross-identifications, when available, generates a pair of positional
residuals:
$$
\left\{
\begin{array}{rcl}
\displaystyle \Delta x_{i,j,k}&\!\!\!=\!\!\!&x_{i,j,k}^{\rm corr}-
(X_i^{\rm master})^{T_{j,k}}\\
\displaystyle \Delta y_{i,j,k}&\!\!\!=\!\!\!&y_{i,j,k}^{\rm corr}-
(Y_i^{\rm master})^{T_{j,k}},\\
\end{array}
\right.
$$
which reflect the residuals in the GD (with the opposite sign), and
depend on where the $i$-star fell on the $k$/$j$ work-image/file (plus
random deviations due to non-perfect PSF-fitting, photon noise, and
errors in the transformations).  [Note that, at the first iteration,
$(x^{\rm corr},y^{\rm corr})_{i,j,k}$$=$$(x,y)_{i,j,k}$.]  In each
chip we have typically 120-130 high-signal stars in common with the
master frame, leading to a total of $\sim$5500 residual pairs per
chip.

These residuals were then collected into a look-up table made up of
11$\times$25 elements, each related to a region of 186.4$\times$184.4
pixels (25$\times$11 elements of 184.4$\times$186.4 for chip [4]).
We chose this particular grid setup because it offers the best
compromise between the need of an adequate number of grid points to
model the GD (the larger, the better) and an adequate sampling of each
grid element (we required to have at least 10 pairs of residuals in
each grid element).
For each grid element, we computed a set of five
3$\sigma$-clipped\footnote{
The clipping procedure is performed as follow:\ first we compute the
median value of the positional residuals of all the stars within a
given grid element $(m,n)$, then we estimate the $\sigma$ as the 68.27
percentile of the distribution around the median.
Outliers for which residuals are larger than 3$\sigma$ are rejected
iteratively.
We note that the process converge after 2--3 iterations, and that most
of the outliers are poorly measured stars, or mismatches, as at the
very first steps the GD could be as large as 20 pixels, and only later
(as the GD improves) these stars are correctly matched.
} quantities:\ $\overline{x}_{m,n,k}$, $\overline{y}_{m,n,k}$,
$\overline{\Delta x}_{m,n,k}$, $\overline{\Delta y}_{m,n,k}$, and
$P_{m,n,k}$; where $\overline{x}_{m,n,k}$ and $\overline{y}_{m,n,k}$
are the average positions of all the stars within the grid element
$(m,n)$ of the $k$-chip, $\overline{\Delta x}_{m,n,k}$ and
$\overline{\Delta y}_{m,n,k}$ are the average residuals, and
$P_{m,n,k}$ is the number of stars that were used to calculate the
previous quantities. These $P_{m,n,k}$ will also serve in associating
a weight to the grid cells when we fit the polynomial coefficients.

\subsubsection{Iterations}
\label{4.5}

To obtain the 16 coefficients describing the two polynomials
($a_{q,k}$ with $q=3,\dots,9$, and $b_{q,k}$ with $q=1,\dots,9$) that
represent our GD solution in each chip, we perform a linear least
square fit of the $N$=$m$$\times$$n$=$11$$\times$25$=$275 cells
(hereafter we will use the notation $p$$=$$1,\dots,N$, instead of the
two $m$$=$$1,\dots,11$ and $n$$=$$1,\dots,25$).  In the linear least
square fit, we can safely consider the errors on the average positions
$\overline{x}_{p,k}$, $\overline{y}_{p,k}$ (i.e.,
$\overline{x}_{m,n,k}$, $\overline{y}_{m,n,k}$) negligible with
respect to the uncertainties on the average residuals
$\overline{\Delta x}_{p,k}$, $\overline{\Delta y}_{p,k}$ (i.e.,
$\overline{\Delta x}_{m,n,k}$, $\overline{\Delta y}_{m,n,k}$).  Thus,
for each chip, we can compute the average distortion correction in
each cell $(\overline{\delta x}_{p,k}$,~$\overline{\delta y}_{p,k})$
with $N$ relations of the form:
$$
k=1,\dots,4; ~  p=1,\dots,N: ~ 
\left\{
\begin{array}{c}
\overline{\delta x}_{p,k}=\displaystyle \sum_{q=3}^9 a_{q,k} t_{q,p,k}\\
\overline{\delta y}_{p,k}=\displaystyle \sum_{q=1}^9 b_{q,k} t_{q,p,k}\\
\end{array}
\right.
$$ 
(where $t_{1,p,k}=\overline{\tilde{x}}_{p,k}$,
$t_{2,p,k}=\overline{\tilde{y}}_{p,k}$, \dots,
$t_{9,p,k}={{\overline{\tilde{y}}}}^3_{p,k}$), concerning the 16
unknown quantities -- or fitting parameters -- $a_{q,k}$ and
$b_{q,k}$, for each chip.

In order to solve for $a_{q,k}$ and $b_{q,k}$, we defined, for each
chip, one 9$\times$9 matrix $\mathcal M_k$ and two 9$\times$1 column
vectors ${\mathcal V}_{a,k}$ and ${\mathcal V}_{b,k}$:
$$
\!\mathcal{M}_k\!=\!\displaystyle  \left(
\begin{array}{cccc}
\sum_p P_{p,k}  t_{1,p,k}^2& \sum_p P_{p,k} t_{1,p,k} t_{2,p,k}  \!&\cdots&\!\! \sum_p P_{p,k}t_{1,p,k}  t_{9,p,k}\!\!\\ 
\!\!\sum_p  P_{p,k} t_{2,p,k}  t_{1,p,k}\!\!&\!\! \sum_p  P_{p,k} t_{2,p,k}^2\!\!&\!\!\cdots\!\!&\!\!  \sum_p P_{p,k}  t_{2,p,k}  t_{9,p,k}\!\!\\ 
\!\vdots&\!\!\vdots&\!\!\ddots&\!\!\vdots\\
\!\!\sum_p P_{p,k} t_{9,p,k} t_{1,p,k} \!\!&\!\! \sum_p P_{p,k} t_{9,p,k} t_{2,p,k}\!&\cdots& \sum_p P_{p,k} t_{9,p,k}^2\\
\end{array}
\right);
$$
$$ 
\mathcal{V}_{a,k}=\left(
\begin{array}{c}
\sum_p P_{p,k} t_{1,p,k} \overline{\Delta x}_{p,k}\\  
\sum_p P_{p,k} t_{2,p,k} \overline{\Delta x}_{p,k}\\ 
\vdots  \\ 
\sum_p P_{p,k} t_{9,p,k} \overline{\Delta x}_{p,k}\\
\end{array}
\right);
\,\,\,\,\,
\mathcal{V}_{b,k}=\left(
\begin{array}{c}
\sum_p P_{p,k} t_{1,p,k} \overline{\Delta y}_{p,k}\\
\sum_p P_{p,k} t_{2,p,k} \overline{\Delta y}_{p,k}\\
\vdots \\
\sum_p P_{p,k} t_{9,p,k} \overline{\Delta y}_{p,k}\\
\end{array}
\right).
$$

\begin{table*}[t!]
\centering
\caption{Coefficients of the third-order polynomial in each chip used
to represent our geometric distortion in the final solution for the
$V$ filter.}
\label{tab:V}
\footnotesize{
\centering
\begin{tabular}{ccrrrrrrrr}
& & & & & & & & & \\ 
\hline
\hline
& & & & & & & & & \\ 
Term $\!(q)\!\!\!\!\!$&Polyn.$\!\!\!\!\!\!\!\!\!$
&$a_{q,[1]}$&$b_{q,[1]}$&$a_{q,[2]}$&$b_{q,[2]}$&$a_{q,[3]}$&$b_{q,[3]}$&$a_{q,[4]}$&$b_{q,[4]}$\\
& & & & & & & & & \\ 
\hline
& & & & & & & & & \\ 
1&$\tilde{x}$&            $ +0.0000 \!$&$\! +6.3877 \!$&$\! +0.0000 \!$&$\! +0.1638 \!$&$\! +0.0000 \!$&$\! -6.2855 \!$&$\! +0.0000 \!$&$\! +1.0239 $\\
2&$\tilde{y}$&            $ +0.0000 \!$&$\!+20.2053 \!$&$\! +0.0000 \!$&$\! -2.7878 \!$&$\! +0.0000 \!$&$\!+20.2021 \!$&$\! +0.0000 \!$&$\!-16.5767 $\\
3&$\tilde{x}^2$&          $ -7.1942 \!$&$\! +0.6063 \!$&$\! -0.1033 \!$&$\! +0.6919 \!$&$\! +7.2181 \!$&$\! +0.6900 \!$&$\! +0.9713 \!$&$\!-15.7873 $\\
4&$\tilde{x}\tilde{y}$&   $ +3.2801 \!$&$\!-10.7004 \!$&$\! +3.6108 \!$&$\! -0.2218 \!$&$\! +3.5260 \!$&$\!+10.5439 \!$&$\!-13.8197 \!$&$\! +0.3543 $\\
5&$\tilde{y}^2$&          $-12.2026 \!$&$\!+11.3321 \!$&$\! -0.3165 \!$&$\!+11.7370 \!$&$\!+12.1294 \!$&$\!+11.7808 \!$&$\! +0.0850 \!$&$\! -9.0966 $\\
6&$\tilde{x}^3$&          $ -1.1416 \!$&$\! -0.0136 \!$&$\! -1.1606 \!$&$\! +0.0168 \!$&$\! -1.0659 \!$&$\! +0.0204 \!$&$\!-13.1945 \!$&$\! +0.0213 $\\
7&$\tilde{x}^2\tilde{y}$& $ -0.1559 \!$&$\! -2.5312 \!$&$\! -0.0204 \!$&$\! -2.5133 \!$&$\! +0.1460 \!$&$\! -2.6083 \!$&$\! +0.1596 \!$&$\! -5.4927 $\\
8&$\tilde{x}\tilde{y}^2$& $ -5.7174 \!$&$\! -0.3752 \!$&$\! -5.7339 \!$&$\! -0.1705 \!$&$\! -5.8841 \!$&$\! +0.2810 \!$&$\! -2.3973 \!$&$\! +0.0324 $\\
9&$\tilde{y}^3$&          $ -0.2740 \!$&$\!-12.6714 \!$&$\! +0.0087 \!$&$\!-12.4949 \!$&$\! +0.1189 \!$&$\!-13.1913 \!$&$\! +0.0876 \!$&$\! -0.9556 $\\
& & & & & & & & & \\ 
\hline
\end{tabular}}
\end{table*}

The solution is given by two 9$\times$1 column vectors $\mathcal A_k$
and $\mathcal B_k$, containing the best fitting values for $a_{q,k}$
and $b_{q,k}$, obtained as:
$$
{\mathcal A_k}=\left(
\begin{array}{c}
a_{1,k}\\
a_{2,k}\\
\vdots\\
a_{9,k}\\
\end{array}
\right)
={\mathcal M_k}^{-1}{\mathcal V}_{a,k};
\,\,\,\,\,\,\,\,\,\,\,\,\,\,\,
{\mathcal B_k}=\left(
\begin{array}{c}
b_{1,k}\\
b_{2,k}\\
\vdots\\
b_{9,k}\\
\end{array}
\right)
={\mathcal M_k}^{-1}{\mathcal V}_{b,k}.
$$

With the first set of calculated coefficients $a_{q,k}$ and $b_{q,k}$
we computed the corrections $\delta x_{i,j,k}$ and $\delta y_{i,j,k}$
to be applied to each $i$-star of the $k$-chip in each $j$-MEF file,
but actually we corrected the positions only by half of the
recommended values, to guarantee convergence.  With the new improved
star positions, we start-over and recalculated new residuals.  The
procedure is iterated until the difference in the average corrections
from one iteration to the following one -- for each grid point --
became smaller than 0.001 pixels. Convergence was typically reached
after $\sim$20--30 iterations.

\subsection{The GD solution}
\label{sec4.6}

Once new corrected star positions have been obtained for all the
images, we can derive a new master frame, and consequently improve our
GD solution for each chip, simply by repeating the procedure used to
determine the polynomial coefficients.  At the end of each iteration,
star positions in the newly derived master frame are closer than before
to the ones of a distortion-free frame, and provide a better reference
on which to calculate the GD correction.  After 15 such iterations,
we were able to reduce star-position residuals from the
initial average of $\sim$4 pixels down to 0.085 pixels ($\sim$20 mas)
(or $\sim$15 mas for each single coordinate).  [A further iteration
proved to give no significant improvements to our solution.]

In Fig.~\ref{fig4} we show -- for each chip -- the residual of
uncorrected star positions versus the predicted positions of our final
master frame, which is representative of our GD solution.  For each
chip, we plot the 11$\times$25 cells used to model the GD, each with
its distortion vector magnified by a factor of 25.  Residual vectors
go from the average position of the stars belonging to each grid cell
$(\overline{x},\overline{y})$ to the corrected one.  We also show the
overall trend of residuals $\Delta x$, $\Delta y$ along $x$ and $y$
directions.  Note the symmetric shape of the geometric distortion
around the center of the FoV.  In Fig.~\ref{fig5} we show, in the same
way, the remaining residuals after our GD solution is applied.  This
time we magnified the distortion vectors by a factor of 500.  [Note
that, close to chip edges, remaining residuals are larger that the
average.  We suggest to exclude those regions for high precision
astrometry.]  The coefficients of the final solution for the four
chips are given in Table~\ref{tab:V}.

\begin{figure*}[t!]
\centering
\includegraphics[width=12cm]{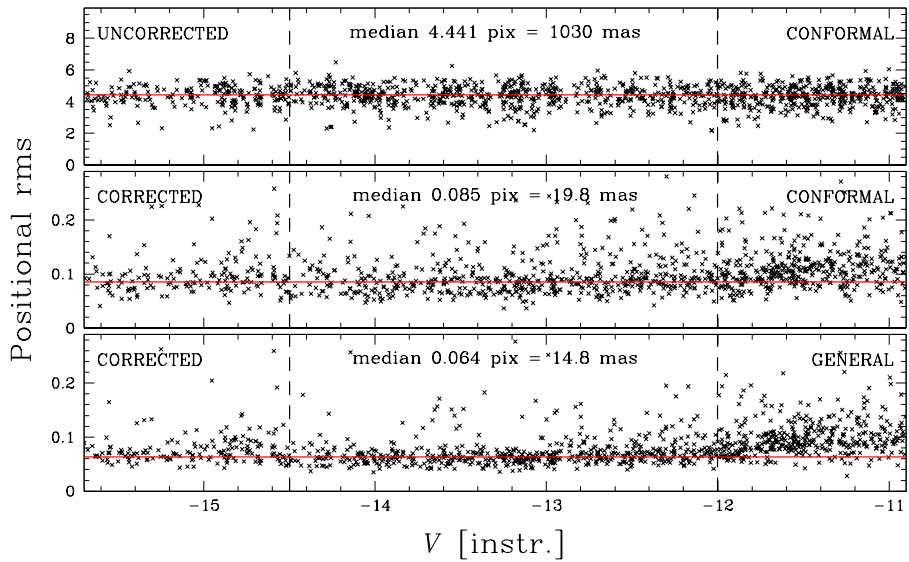}
\caption{(\textit{Top}): Positional r.m.s.\ as a function of the
  instrumental $V$ magnitude without GD correction.
  (\textit{Middle}): Same plot, but with GD correction.  Corrected
  catalogs are transformed into the reference frame using a conformal
  transformation.  (\textit{Bottom}): Same plot, but using the most
  general linear transformations (6 parameters) to bring the corrected
  catalogs into the reference frame, instead of a conformal
  transformation (4 parameters). Vertical dashed lines mark the
  magnitude interval used to calculate median values for the
  positional r.m.s.}
\label{fig6:Vresiduals}
\vskip 2mm
\end{figure*}

\subsection{Accuracy of the GD Solution}

The best estimate of the true errors in our GD solution is given by
the size of the r.m.s.\ of the position residuals observed in each
work-image, which have been GD-corrected, and transformed into the
reference frame $(X_{i,j,k},Y_{i,j,k})$.  Since each star has been
observed in several work-images and in different regions of the
detectors, the consistency of these star positions, once transformed
in the coordinate system of the distortion-free reference frame
$(X_i^{\rm master}, Y_i^{\rm master})$, immediately quantifies how
well we are able to put each image into a distortion-free system.

\begin{figure*}[ht!]
\centering
\includegraphics[width=12cm]{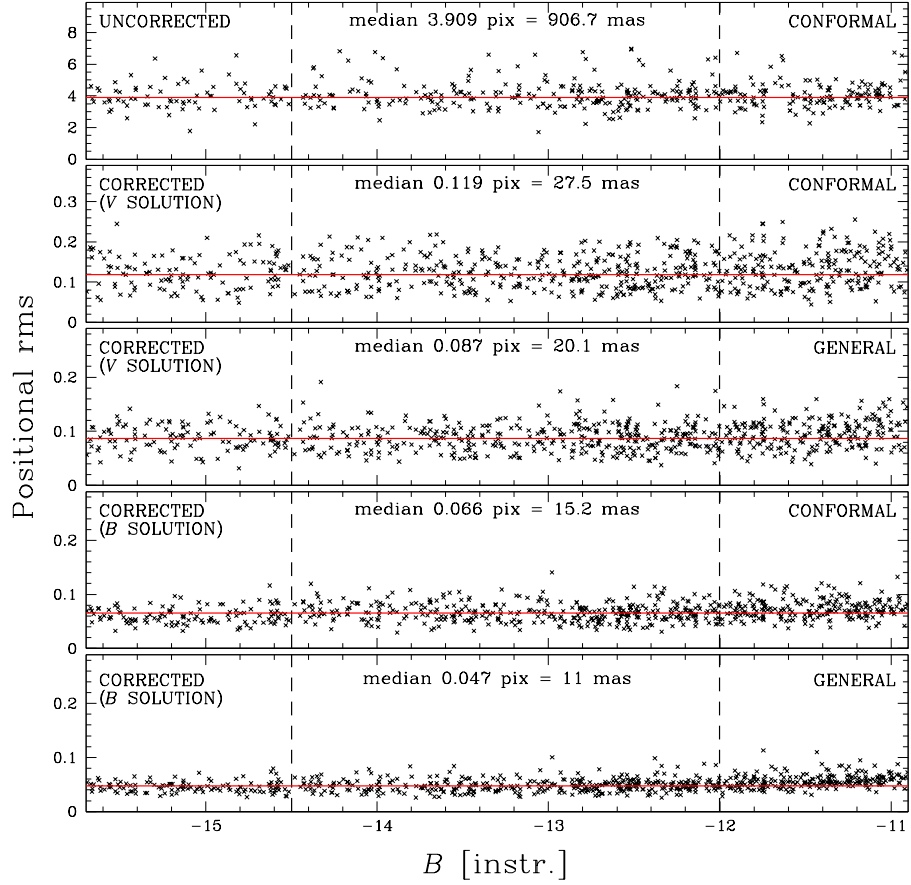}
\caption{ (\textit{From Top to Bottom}): Positional r.m.s.\ as
function of the instrumental $B$ magnitude when no GD correction is
applied at all.  (\textit{Next}): Same plot, but after the $V$-derived
GD correction applied, and using conformal transformations to
transform corrected catalogs into the reference frame.
(\textit{Next}): Same plot, but using the most general linear
transformations (6 parameters).  (\textit{Next}): Positional r.m.s.\
after the GD correction obtained from $B$ images is applied, and using
conformal transformations.  (\textit{Bottom}): Same as above but using
a general linear transformations.  Vertical dashed lines mark the
magnitude interval used to calculate median values for the positional
r.m.s.}
\label{fig7:Bresiduals}
\vskip 2mm
\end{figure*}

In the top panel of Fig.\ \ref{fig6:Vresiduals} we show the size of
these r.m.s.\ versus the instrumental $V$ magnitude, before GD
correction is applied -- {\it at all} -- to the observed positions,
before transforming them into the master-frame using a conformal
transformation.  The instrumental magnitude ($m_{\rm instr.}$) has
been computed as the sum of the pixel's digital numbers (DNs) under
the best fitted PSF (i.e.\ $m_{\rm instr.}$$=$$-2.5\log{[\Sigma ({\rm
DNs}})]$).  For reference, in images with a seeing of $0\farcs6$,
saturation begins at $m_{\rm instr.}$$=$$-$14.1, while if the seeing
is $\sim$$1\farcs0$, the saturation level can reach $m_{\rm
instr.}$$=$$-$15.2.  (This simply means that in these two cases, the
brightest pixels contain $\sim$12\% or $\sim$4\% of the flux,
respectively, enabling to collect more or fewer photons before
saturation is reached in the brightest pixel.)

The r.m.s.\ are computed from the values $\sqrt{[(X_{i}^{\rm
master})^{T_{j,k}}-X_i^{\rm master}]^2 + [(Y_{i}^{\rm
master})^{T_{j,k}}-Y_i^{\rm master}]^2 }$.  Only stars in the
master-list observed in at least 9 images, and within $\sim$2.5
magnitudes below the saturation level (between the dashed lines) were
considered to test the accuracy of the GD solution, because faint
stars are dominated by random errors. Note, however, that we applied
our GD solution to all the sources in our catalogs.  We can see that
if no GD correction is applied, the positional r.m.s.\ exceed 4.4
pixels (i.e.\ a whole arcsec).  In some locations on the chips
individual displacements can exceed 20 pixels (5 arcsec), see
Fig.~\ref{fig4}.

Middle panel of Fig.\ \ref{fig6:Vresiduals} shows that, once our GD
correction is applied, the positional r.m.s.\ reach an accuracy of
$\sim$20 mas for high signal-to-noise ratio (S/N) stars.  It is worth
noting that saturated stars ($V_{\rm instr.}$$<$$-$14.5) are also
reasonably well measured.  When a 6-parameter linear transformation
(the most general possible linear transformation, hereafter simply
{\it general} transformation) is applied, most of the residuals
introduced by variation of the telescope+optics system (due to thermal
or gravity-induced flexure variation, and/or differential atmospheric
refraction) are absorbed, and the r.m.s.\ further reduces to 0.064
pixels ($\sim$15 mas, see bottom panel of Fig.\
\ref{fig6:Vresiduals}).  Note that when at least a dozen of high S/N
stars are present in the field, this kind of transformation should
always be preferred for relative astrometry.  The corners of the FoV,
however, show systematic residuals larger than the r.m.s.\ (see also
Fig.\ \ref{fig4} and \ref{fig5}), indicating problems of stability of
the geometric distortion solution over the 6-day period of
observations.

If the stellar density in the field is high enough, and if relative
astrometry is the goal of the investigation, these residual systematic
errors could be further reduced with a local transformation approach
(Bedin et al.\ \cite{bedin03}; Paper~I, II, and in Bellini et al.\
\cite{bellini09}, hereafter Paper~III).

\begin{table*}[t!]
\caption{Our distortion coefficients for the $B$ filter. }
\label{tab:B}
\centering
\footnotesize{
\begin{tabular}{ccrrrrrrrr}
& & & & & & & & & \\ 
\hline
\hline
& & & & & & & & & \\ 
Term $\!(q)\!\!\!\!\!$&Polyn.$\!\!\!\!\!\!\!\!\!$
&$a_{q,[1]}$&$b_{q,[1]}$&$a_{q,[2]}$&$b_{q,[2]}$&$a_{q,[3]}$&$b_{q,[3]}$&$a_{q,[4]}$&$b_{q,[4]}$\\
& & & & & & & & & \\ 
\hline
& & & & & & & & & \\ 
1&$\tilde{x}$&            $ +0.0000 \!$&$\! +6.3397 \!$&$\! +0.0000 \!$&$\! +0.0040 \!$&$\! +0.0000 \!$&$\! -5.9624 \!$&$\! +0.0000 \!$&$\! +1.2625 $\\
2&$\tilde{y}$&            $ +0.0000 \!$&$\!+20.8097 \!$&$\! +0.0000 \!$&$\! -2.5055 \!$&$\! +0.0000 \!$&$\!+20.3966 \!$&$\! +0.0000 \!$&$\!-16.5971 $\\
3&$\tilde{x}^2$&          $ -7.1860 \!$&$\! +0.5848 \!$&$\! -0.0755 \!$&$\! +0.6457 \!$&$\! +7.1015 \!$&$\! +0.6013 \!$&$\! +1.0365 \!$&$\!-15.8413 $\\
4&$\tilde{x}\tilde{y}$&   $ +3.4380 \!$&$\!-10.9259 \!$&$\! +3.6067 \!$&$\! -0.2598 \!$&$\! +3.4450 \!$&$\!+10.5167 \!$&$\!-13.7325 \!$&$\! +0.3114 $\\
5&$\tilde{y}^2$&          $-12.1617 \!$&$\!+11.4846 \!$&$\! +0.0393 \!$&$\!+11.6489 \!$&$\!+12.4396 \!$&$\!+11.4898 \!$&$\! +0.0851 \!$&$\! -8.9071 $\\
6&$\tilde{x}^3$&          $ -1.0551 \!$&$\! -0.0383 \!$&$\! -1.2039 \!$&$\! +0.0094 \!$&$\! -1.0662 \!$&$\! +0.0081 \!$&$\!-13.0938 \!$&$\! -0.0209 $\\
7&$\tilde{x}^2\tilde{y}$& $ -0.1953 \!$&$\! -2.4242 \!$&$\! +0.0102 \!$&$\! -2.6102 \!$&$\! +0.0762 \!$&$\! -2.5362 \!$&$\! +0.0593 \!$&$\! -5.3879 $\\
8&$\tilde{x}\tilde{y}^2$& $ -5.6316 \!$&$\! -0.3257 \!$&$\! -5.9606 \!$&$\! +0.0429 \!$&$\! -5.7653 \!$&$\! +0.2457 \!$&$\! -2.3617 \!$&$\! -0.0096 $\\
9&$\tilde{y}^3$&          $ -0.3043 \!$&$\!-13.2670 \!$&$\! -0.0894 \!$&$\!-13.3468 \!$&$\! +0.0682 \!$&$\!-13.3496 \!$&$\! +0.0351 \!$&$\! -0.9604 $\\
& & & & & & & & & \\ 
\hline
\end{tabular}}
\end{table*}

\subsection{GD correction for the $B$ filter}
\label{sec6}

Every LBC-Blue filter constitutes a different optical element which
could slightly change the optical path and introduce -- at some
level -- changes in the GDs.  To test the filter-dependency of our GD
solution derived for the $V$ filter, we corrected the positions measured
on each $B$ images with our $V$-filter-derived GD solution and studied
the positional r.m.s.

Analogously to Fig.\ \ref{fig6:Vresiduals}, we show in the top panel
of Fig.\ \ref{fig7:Bresiduals} the positional r.m.s.\ as a function of
the instrumental $B$ magnitude when no GD correction is applied to the
observed positions, and where conformal transformations were used to
bring each catalog into the reference frame.  In the following second panel
we show the positions corrected with the GD-solution obtained from $V$
images, again using conformal transformations.  In the third
panel, we show the same r.m.s.\ once the corrected positions are
transformed with a general (linear) transformation.

Since we found these r.m.s. significantly larger ($>$20 mas) than the
ones obtained for the $V$ filter, we decided to independently solve
for the GD also for the $B$ images.  We repeated the procedure
described in the previous sections, but this time using our $V$ filter
GD correction as a first guess.  Table~\ref{tab:B} contains the
coefficients derived for our GD solution using only images in the $B$
filter. The values of the coefficients are consistent with those
obtained for the $V$ filter, but different at a level of few percent.

In the fourth panel of Fig.\ \ref{fig7:Bresiduals} we show that the
positional r.m.s.\ (now corrected with the $B$-derived GD solution and
conformally transformed into the reference frame) are significantly
smaller, down to $\sim0.07$ pixels.  Finally, a general linear
transformation further reduces these values to less than $\sim$0.05
pixels, i.e.\ $\sim$11 mas ($\sim$8 mas in each coordinate, see bottom
panel of Fig.\ \ref{fig7:Bresiduals}).

It might seem that the GD solution derived from images collected with
the $B$ filter is even better than the one derived from the $V$ one,
but that would be a wrong interpretation. Indeed, these smaller r.m.s.
are due to the fact that the chip inter-comparison is not complete,
having at our disposal only small dithers for the $B$ filter.

\begin{table}[t!]
\centering
\caption{Inter-chip transformation parameters, with formal errors.
  The absolute $x$-scale factor of chip [2] in its reference pixel is
  229.7$\pm$0.1 mas.  The values for $\theta$ are expressed in
  degrees.}
\label{tab:param}
\begin{tabular}{ccccc}
 & & & & \\
\hline
\hline
 & & & & \\
Parameter &       $k=$[1] & $k=$[2] & $k=$[3] &  $k=$[4]\\
 & & & & \\
\hline
 & & & & \\
$\alpha_{k}/\alpha_{[2]}$&1.01482&1.00000&1.01445&1.0073\\
                &$\pm$0.00006& &$\pm$0.00006& $\pm$0.0001\\
 & & & & \\
$\theta_{k}$$-$$\theta_{[2]}$&$-$0.175&0.000&0.198&0.005\\
                &$\pm$0.003&&$\pm$0.04&$\pm$0.003\\
& & & & \\
$(x_{[2]}^{\rm corr})_k$&3135.0&1025.00&$-$1088.1&948.36\\
                &$\pm$0.1& &$\pm$0.1 &$\pm$0.08\\
& & & & \\
$(y_{[2]}^{\rm corr})_k$&2311.2& 2305.00&2307.4& 5684.9\\
                &$\pm$0.1& &$\pm$0.1&$\pm$0.2\\
 & & & & \\
\hline
\end{tabular}
\end{table}

\begin{figure*}[t!]
\centering
\includegraphics[width=18.4cm]{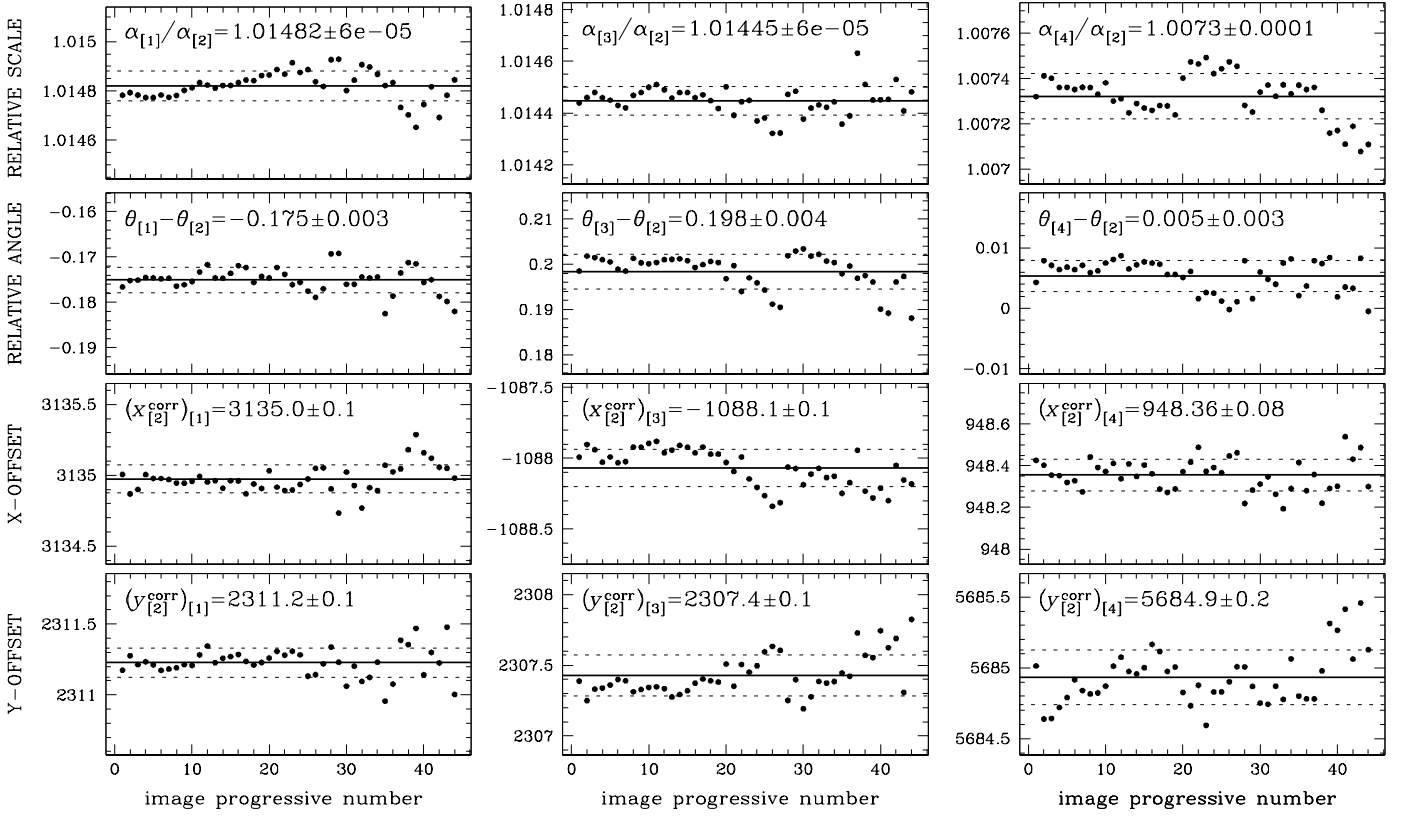}
\caption{ For all the 44 $V$ images (sorted by chronological order) we
show the variation of the linear quantities of chips [1], [3], and
[4], relative to those of chip [2].  From top to bottom: {\it (i)} the
relative scale $\alpha_{k}/\alpha_{[2]}$; {\it (ii)} the relative
position angle $\theta_{k}$$-$$\theta_{[2]}$, in degrees; {\it (iii)}
the offset $(x_{[2]}^{\rm corr})_k$
of the central pixel of chips [1], [3], [4] with respect to the
corrected pixel-coordinate system of chip [2];
and {\it (iv)} the same for the offset $(y_{[2]}^{\rm corr})_k$.
Images from number 20 to 44 are those affected by an anomalously high
background value, and present larger scatters.  Horizontal dot-dashed
lines show the mean values, while short-dashed lines mark $\pm
1\sigma$ (r.m.s.).}
\label{fig8}
\end{figure*}

\section{Relative positions of the chips}
\label{sec:8}

Now that we are able to correct each of the four catalogs (one
per chip) of every LBC image for GD, we want to put them into a common
distortion-free system.  This can be done in a way conceptually very
similar to the one used to solve for the GD within each chip.  

We could then simply conformally transform the corrected positions of
chip $k$ into the distortion-corrected positions of
chip~[2], using the following relations:%
\footnote{Chip [2] occupies a central position within the
LBC-Blue layout (see Fig.~\ref{fig1}), therefore we chose to adopt it
as the reference chip with respect to which we compute relative
scales, orientations, and shifts of the other chips.}
$$
\displaystyle
\begin{array}{rcl}
\left(
\begin{array}{c}
x^{\rm corr}_{[2]}\\
y^{\rm corr}_{[2]}\\
\end{array}
\right)
&\!=\!& \frac{\alpha_{[2]}}{\alpha_{k}}\!
\left[
\begin{array}{cc}
\!\! \cos(\theta_{[2]}-\theta_{k}) & \sin(\theta_{[2]}-\theta_{k}) \\
\!\!-\sin(\theta_{[2]}-\theta_{k}) & \cos(\theta_{[2]}-\theta_{k}) \\
\end{array}
\!\right]
\!\left(
\begin{array}{c}
x_{k}^{\rm corr} - (x_{\circ})_k \\
y_{k}^{\rm corr} - (y_{\circ})_k \\
\end{array}
\!\right)
+
\\
&\!+\!&\!\!\!\!
\begin{array}{ll}
&
\left(
\begin{array}{c}
\!(x_{[2]}^{\rm corr})_k\\
\!(y_{[2]}^{\rm corr})_k\\
\end{array}
\!\right);
\end{array}
\end{array}
$$ 
where -- following the formalism in AK03 -- we indicate the scale
factor as $\alpha_{k}$, the orientation angle as $\theta_{k}$, and the
positions of the center of the chip $(x_\circ,y_\circ)_k$ in the
corrected reference system of chip [2] as $(x_{[2]}^{\rm corr})_k$ and
$(y_{[2]}^{\rm corr})_k$.  Of course, for $k$=2, corrected and
uncorrected values of $(x_\circ,y_\circ)_{[2]}$ are identical.  The
values of the interchip transformation parameters are given in
Table~\ref{tab:param}.

In Figure~\ref{fig8} we show our calculated quantities for chip [1],
[3], and [4], relative to chip [2], using all $V$-images (numbered
from 1 to 44, in chronological order).  Top panels show all the values
for the relative scale $\alpha_k/\alpha_{[2]}$.  The panels in the
second row show the variations of the relative angle
$\theta_k$$-$$\theta_{[2]}$, while the panels in the third and fourth
row show the relative offsets $(x_{[2]}^{\rm corr})_k$ and
$(y_{[2]}^{\rm corr})_k$, respectively.  The mean values of
$\alpha_k/\alpha_{[2]}$, $\theta_k$$-$$\theta_{[2]}$, $(x_{[2]}^{\rm
corr})_k$, and $(y_{[2]}^{\rm corr})_k$, are collected in
Table~\ref{tab:param}.

The differences in scale observed among the chips merely reflect the
different distances of the respective (arbitrarily adopted) reference
pixels from the principal axes of the optical system, roughly at the
center of the LBC-Blue FoV (see Fig.~\ref{fig1}). This is also the
reason why the values of the relative scales for [1] and [3] are
similar.

Finally, we inter-compared star positions in the Digital Sky Surveys
with those of our reference frame, and derived an absolute $x$-scale
factor for chip [2] in its reference point $(x_\circ,y_\circ)_{[2]}$.
We found a value for $\alpha_{[2]}$$=$ 229.7$\pm$0.1 mas ($\equiv$1 pixel
on the LBC-Blue chip $[2]$); the error reflects the scale stability
under the limited conditions explored (see next Section).


\begin{figure}[t!]
\centering
\includegraphics[width=6.5cm]{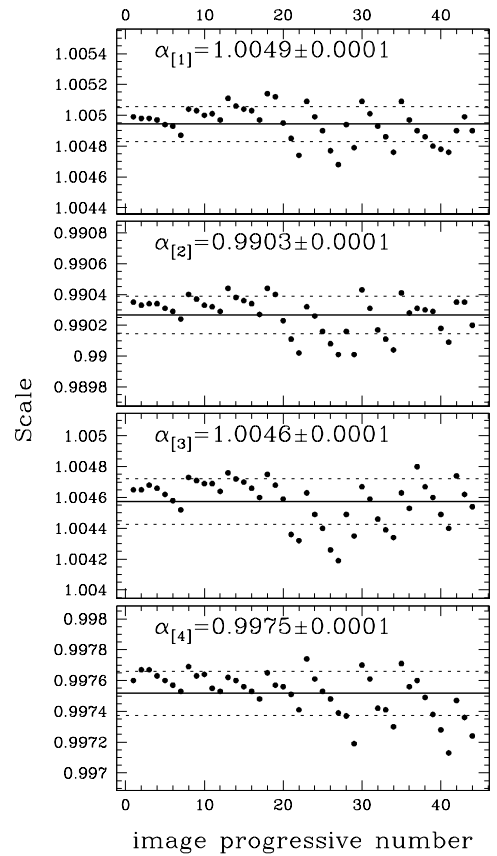}
\caption{Scale factor $\alpha_k$ with formal errors, relative to the
scale of the master frame ($0\farcs2319$) [note, not $\alpha_{[2]}$],
for all the 44 $V$ images (in chronological order).  Solid lines mark
the average values, while, dashed lines mark $\pm$1$\sigma$ (r.m.s.).}
\label{fig9}
\end{figure}

As a further test on our GD-correction solution (and its
utility for a broader community), we reduced two dithered images with
independent, commonly-used software ({\sf{DAOPHOT}}, Stetson 1987) and
applied (step by step) the procedure given in the previous Sections to
the obtained raw-pixel coordinates. We verified that our solution is
able to bring the two images (four chips each) into a common
distortion-free system with an average error $\lesssim20$ mas, i.\ e.\
within the positioning single-star error of an independent code.

\section{Stability of the solution}
\label{sec:9}

In this section we explore the stability of our derived GD solution on
the limited time baseline and condition samplings offered by our
observations.

Table~\ref{tab1} shows us that for $V$ images we can explore only a
time baseline of the order of an hour, and at two different epochs
separated by roughly a week.  Moreover, we have already described in
Sect.~\ref{sec6} how $B$ images provide a somewhat different
GD correction with respect to the $V$-derived GD solution.  It has to
be noted, however, that the $V$-derived GD solution is obtained from
data collected $\sim$2 weeks before the $B$-filter one, therefore we
can not assess if the observed dependencies of the GD solution on the
filter are really due to an effective influence of a different element
in the optical path, or to a filter-independent temporal variation of
the GD.

In Figure \ref{fig9} we show the variation of the individual
(corrected) work-image scale $\alpha_k$, with respect to the master
frame (note that here the reference scale is the one of the master
frame, by definition identically equal to 1, and not the one of
chip[2]), as a function of the progressive image number.  Scale-values
show fluctuations with amplitudes up to 5 parts in $10\,000$, even
within the same night (although the run lasted only about an hour).
%
We also note a clear path of about five consecutive exposures within
each observing block (OB). Indeed, every OB was meant not to last for
more than $\sim$20 minutes, after which the focus of the telescope
needs to be readjusted (and therefore the scale changes). [This is
totally expected for a prime-focus camera with such a short focal
ratio and large FoV; as different pointings cause different
gravity-induced flexures of the large LBT+LBC structure.]
Solid lines mark the average values, while, dashed lines mark
$\pm$1$\sigma$ (r.m.s.).  This seems to suggest that positional
astrometry -- {\it which completely relies on our GD solution} --
could have systematics as large as 250 mas ($\sim$1 pixel) within a
given chip, or up to $0\farcs5$ ($\sim$2 pixels) in the meta-chip
system, although it could be even worse because of the limited
observing conditions explored.  At any rate, one should never rely on
the absolute values of the linear terms provided by our GD corrections
for precise \emph{absolute} astrometry (more in the conclusions).

\begin{figure*}[t!]
\centering
\includegraphics[width=14cm]{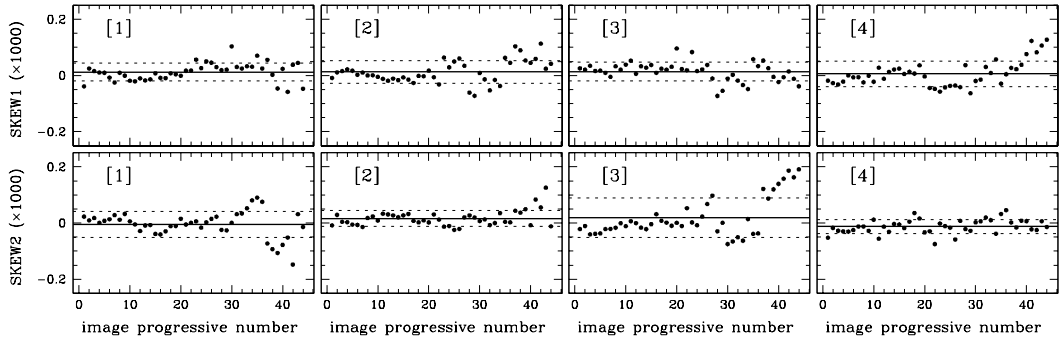}
\caption{As in Fig.~\ref{fig9}, but this time for the observed 
variations in SKEW1 and SKEW2, magnified by a factor of 1000 (see text).}
\label{fig10}
\end{figure*}

Next, we explore the variations of the skew terms: SKEW1, and SKEW2:
SKEW1 indicate whether or not there is a lack of perpendicularity 
between axes, while SKEW2 gives information about the scale differences 
along the two directions. 
In this work, these quantities are defined for each $k$-chip as:\ 
$$  
\begin{array}{c}
{\rm SKEW1}_k  =  (A_k-D_k) / (2\alpha_k)  \\
{\rm SKEW2}_k  =  (B_k+C_k) / (2\alpha_k), \\  
\end{array}
$$
where $A_k,B_k,C_k,D_k,X_{\circ,k},Y_{\circ,k}$ are the values of a
general 6-parameter linear transformation of the form:
$$ \left\{ 
\begin{array}{c}
X^{\rm master} = A_k x^{\rm corr}_k + B_k y^{\rm corr}_k + X_{\circ,k} \\ 
Y^{\rm master} = C_k x^{\rm corr}_k + D_k y^{\rm corr}_k + Y_{\circ,k}. \\
\end{array}
\right.
$$
In Figure~\ref{fig10} we show, for each different chip, the variation
of SKEW1 and SKEW2 parameters (magnified by a factor of 1000).  

As expected (because compared with their average, i.e.\ the master
frame), the average values of the two skew terms are consistent with
zero, although they show some significant well defined trend with
time.  [For example, images with progressive number from 20 to 44
(those affected by the anomalously high background, Feb.~27$^{\rm
th}$), show a trend and a larger scatter with respect to the previous
ones (Feb.~22$^{\rm nd}$)].  Solid lines mark the average values,
while, dashed lines mark $\pm$1$\sigma$ (r.m.s.).

\begin{figure*}[t!]
\centering 
\includegraphics[width=8cm]{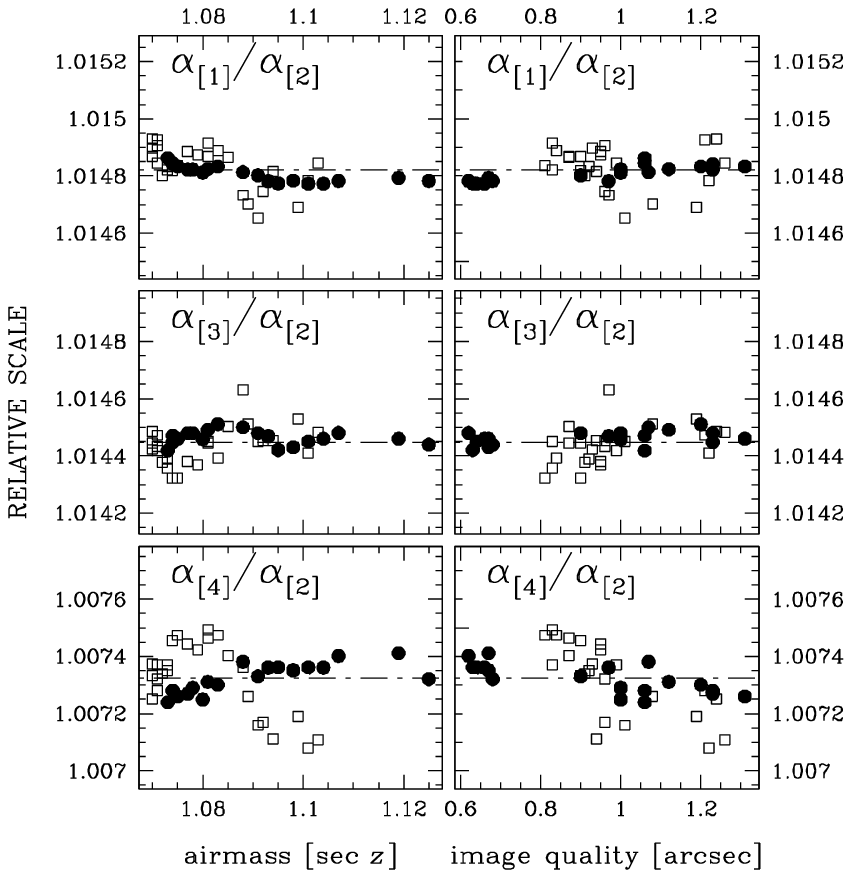}\hskip 2mm
\includegraphics[width=8cm]{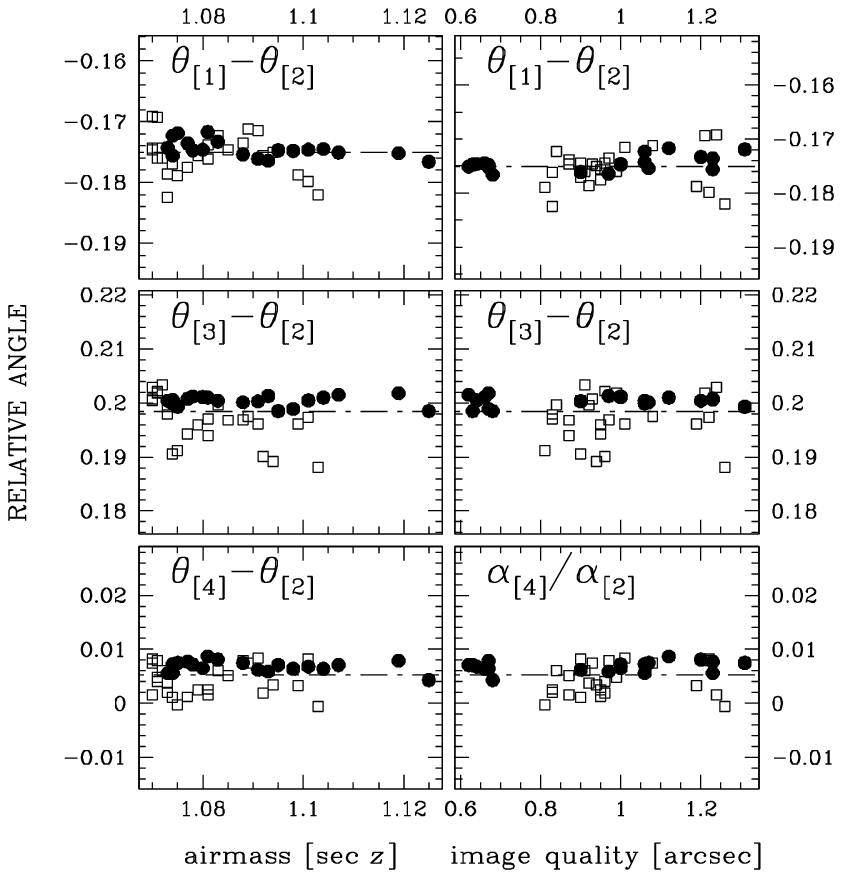}\\
\caption{ \textit{(Left:)} Variation of the relative scale
$\alpha_k/\alpha_{[2]}$ with respect to airmass and image quality.
Full circles are images obtained on Feb.~22$^{\rm nd}$, while open
squares are those of Feb.~27$^{\rm th}$.  \textit{(Right:)} The same, but
for the relative angle $\theta_k$$-$$\theta_{[2]}$, in degrees. Horizontal lines
mark the mean values.}
\label{fig11}
\end{figure*}

It is interesting to check -- at this point -- if the observing parameters
correlate, or not, with temporal variations of the measured inter-chip
transformation parameters.  Figure~\ref{fig11}, shows the variation of
$\alpha_k/\alpha_{[2]}$ (left panels) and $\theta_k$$-$$\theta_{[2]}$
(right panels) with respect to airmass and image quality.  Full
circles mark images obtained on Feb.  22$^{\rm nd}$, while open
squares are those of Feb.  27$^{\rm th}$ (affected by high background
values).  The relative scale $\theta_k$$-$$\theta_{[2]}$, and the
relative angle $\alpha_k/\alpha_{[2]}$ both present larger scatter in
observations collected on Feb.  27$^{\rm th}$, than those of
Feb.~22$^{\rm nd}$. Again, solid lines mark the average values, while,
dashed lines mark $\pm$1$\sigma$ (r.m.s.).

\section{Conclusions}
\label{sec:10}

By using a large number of well dithered exposures we have found a set
of third-order-correction coefficients for the geometric distortion
solution of each chip of the LBC-Blue, at the prime focus of the LBT.

The use of these corrections removes the distortion over the entire
area of each chip to an accuracy of $\sim$0.09 pixel (i.e.\ $\sim$20
mas), the largest systematics being located in the 200-400 pixels
closest to the boundaries of the detectors.  Therefore, we advise the
use of the inner parts of the detectors for high-precision astrometry.
The limitation that has prevented us from removing the distortion at
even higher level of accuracies -- in addition to atmospheric effects
and to the relatively sparsity of the studied field -- is the
dependency of the distortion on the scale changes that result from
thermal and/or gravitational induced variations of the
telescope+optical structure.

If a dozen (or more) well distributed high S/N stars are available
within the same chip, a general 6-parameter linear transformation
could register relative positions in different images down to about 15
mas.  If the field is even more densely populated, then a local
transformation approach [as the one adopted in Bedin et al.\
(\cite{bedin03}), from space, or in Paper~I, II, III) from ground] can
further reduce these precisions to the mas level.  [Indeed, using
these techniques and this very same data-set we were able to reach a
final precision of $\sim$1 mas$\,$yr$^{-1}$ (Bellini et al., submitted
to \aap\ Letters)].

These are the precisions and accuracies with which we can hope to
bring one image into another image by adopting:\ {\it conformal}, {\it
general}, or {\it local} transformations.  In the case of absolute
astrometry, however, the accuracies are much lower.  During the
available limited number of nights of observations (and atmospheric
conditions), we observed scale-variations up to 5 parts in $10\,000$,
even during the same night.  This implies that astrometric accuracy
-- which completely relies on our GD solution -- can not be better
than $\sim$250 mas ($\sim$1 pixel) within a given chip (from center to
corners), and can be as large as $0\farcs5$ ($\sim$2 pixels) in the
meta-chip system.  This value is in-line with the meta-chip stability
observed in other ground-based WFI (Paper~I), and absolutely excellent
for a ground-based prime-focus instrument with such a small focal
ratio and large FoV.

Thankfully, several stars from astrometric catalogs such as the
UCAC-2, GSC-2, 2MASS, will be always available within any given
LBC-Blue large FoV. 
These stars, in addition to provide a link to absolute astrometry (as
done for example in Rovilos et al.\ 2009), will enable constrains of
linear terms in our GD solution, and to potentially reach an absolute
astrometric precision of 20 mas.
The fact that we are able to reach good astrometric precision also for
saturated stars will make the comparison between these catalogs and
the sources measured in the -- generally deeper -- LBC images, even
easier.

For the future, more data and a longer time-baseline are needed to
better characterize the GD stability of LBC@LBT detectors on the
medium and long time term. This could make it possible to:\ (1)
determine a multi-layer model of the distortion which would properly
disentangle the contributions given by optical field-angle distortion,
light-path deviations caused by filters and windows, non-flat CCDs,
CCDs artifacts, alignment errors of the CCD on the focal plane, etc.;
and (2) allow for time-dependent and/or mis-alignments of mirrors,
filters/windows, and CCDs.

\acknowledgements A.B.\ acknowledges support by the CA.RI.PA.RO.\
foundation, and by the STScI under the {\em ``2008 graduate research
assistantship''} program. We warmly thank our friend Alceste Z.\
Bonanos for a careful polishing of the manuscript, and Jay Anderson 
for many useful discussions.



\end{document}